\documentclass[usenatbib, usegraphicx]{mn2e}

\newlength{\colwidth}
\newcommand{\yr}{{\rm yr}} 
\newcommand{\cm}{{\rm cm}}

\newcommand{\kms}{{\rm km}\,{\rm s}^{-1}}
\newcommand{\K}{{\rm K}}

\newcommand{\hkpc}{h^{-1}\,{\rm kpc}}

\newcommand{\erg}{{\rm erg}}
\newcommand{\hMpc}{h^{-1}\,{\rm Mpc}}
\newcommand{\Msun}{{{\rm M}_\odot}}
\newcommand{\hMsun}{{h^{-1}\,{\rm M}_\odot}}

\newcommand{\ion}[2]{\hbox{{\rm #1}\,{\sc #2}}}

\newcommand{\OVI}{\ion{O}{vi}}

\title[The enrichment history of cosmic metals]{The enrichment history of cosmic metals}
\author[R.P.C. Wiersma et al.]{Robert~P.~C.~Wiersma,$^{1,2}$\thanks{E-mail:
wiersma@strw.leidenuniv.nl} Joop~Schaye,$^1$ Claudio~Dalla~Vecchia,$^{1,3}$ C.~M.~Booth,$^1$ \newauthor Tom~Theuns$^{4,5}$ and Anthony~Aguirre$^6$\\ 
$^1$Leiden Observatory, Leiden
  University, P.O. Box 9513, 2300 RA Leiden, The Netherlands\\
$^2$Max-Planck-Institut f\"{u}r Astrophysik, Karl-Schwarzschild-Stra{\ss}e 1, D-85478, Garching, Germany\\
$^3$Max-Planck-Institut f\"{u}r Extraterrestrische Physik, Giessenbachstra{\ss}e, D-85478, Garching, Germany\\
$^4$Institute for Computational Cosmology, Department of Physics,
University of Durham, South Road, Durham, DH1 3LE, UK\\
$^5$Department of Physics, University of Antwerp, Groenenborgerlaan
171, B-2020 Antwerpen, Belgium\\
$^6$SCIPP, University of California, Santa Cruz, CA 95064, USA
}

\begin{document}

\maketitle

\begin{abstract}
We use a suite of cosmological, hydrodynamical simulations to investigate the chemical enrichment history of the Universe. Specifically, we trace the origin of the metals back in time to investigate when various gas phases were enriched and by what halo masses. We find that the age of the metals decreases strongly with the density of the gas in which they end up. 
At least half of the metals that reside in the diffuse intergalactic medium (IGM) at redshift zero (two) were ejected from galaxies above redshift two (three). The mass of the haloes that last contained the metals increases rapidly with the gas density. More than half of the mass in intergalactic metals was ejected by haloes with total masses less than $10^{11}\,\Msun$ and stellar masses less than $10^9\,\Msun$. The range of halo masses that contributes to the enrichment is wider for the hotter part of the IGM. By combining the `when' and `by what' aspects of the enrichment history, we show that metals residing in lower density gas were typically ejected earlier and by lower mass haloes.
\end{abstract}


\section{Introduction}

Despite the fact that `metals' - heavy elements produced by stars -
make up a very tiny portion of the cosmic matter budget, they are of
critical importance to our understanding of the Universe. From a
diagnostic point of view, they are very useful in constraining star
formation and tracing feedback from massive stars and
supernovae. More importantly, however, metals impact the rate at which
gas cools, affecting structure formation on scales of galaxy clusters
down to dust grains. This effect should not go understated since, depending on the temperature, metals can reduce the cooling time by over an order of magnitude for solar metallicity \cite[e.g.][]{Sutherland1993, Gnat2007, Wiersma2009a}.

Observations clearly show that a substantial fraction of the diffuse intergalactic medium (IGM) has been enriched with heavy elements \citep[e.g.][]{Cowie1998,Schaye2000a,Ellison2000,Schaye2003,Simcoe2004,Aracil2004,Schaye2007,Aguirre2008,Danforth2008,Cooksey2010,Pieri2010}. However, the physical mechanism, the timing and the sources of the enrichment all remain unknown. The IGM may have been enriched by the first generations of stars and galaxies at very high redshifts, or it may have been polluted by more massive galaxies at intermediate redshifts. Metals may be carried out into intergalactic space by galactic winds driven by supernovae,
radiation pressure from star light, or by active galactic nuclei
(AGN) \citep[e.g.][]{Cen1999a,Aguirre2001,Aguirre2001z3,Madau2001,Theuns2002,Tornatore2004,Aguirre2005,Oppenheimer2006,Dave2007,Oppenheimer2008,Scannapieco2006b,Kawata2007,Kobayashi2007,Sommer-Larsen2008,Wiersma2009b,Shen2009,Tornatore2010,Choi2010}. In dense environments such as groups and clusters
of galaxies metals may also be efficiently dispersed via processes such as ram pressure stripping or tidal stripping (see e.g.\ \citealt{Borgani2008} for a review).

The most
straightforward way to investigate when the IGM received its metals,
is to plot the metallicity or metal mass as a function of redshift, as
we did for a series of cosmological simulations in \citet{Wiersma2009b}. By comparing models with different physical 
implementations,  one can obtain information about the processes that
are important for IGM enrichment. However, the metals that reside in
the IGM at high redshift may not be the same metals that are in the
IGM at the present day. What is hot IGM gas at high
redshift, may have cooled by low redshift. Intergalactic metals may
also accrete onto galaxies and end up in the ISM or in stars. 

Here we also use cosmological simulations to study the enrichment of the cosmos - noting that the IGM is of particular interest - but we take a novel approach in that we classify
the gas based on its state at the redshift of interest (we choose
$z=0$ and $z=2$) and trace its enrichment history back in time with
the aim of understanding how the observed metals were
put into place. Specifically, we will address the following two
questions: When did gas that resides in the IGM at $z=0$ and $z=2$
receive its metals? And what are the masses of the haloes that last contained these metals?

This paper is organised as follows. In Section~\ref{sec-method}
we describe 
the simulations used, emphasising the details that are
particularly pertinent here. Section~\ref{sec-when} discusses
\textit{when} the gas was enriched, taking each sub-phase in turn.
In Section~\ref{sec-what} we investigate \textit{what} enriched the
gas, using a 
halo-finder to determine what kinds of objects are mainly responsible
for the enrichment of various gas phases. We connect the two
ideas in Section~\ref{sec-conn}, 
relating enrichment time with the last halo that contained
the metals that end up in the IGM. Finally, we summarise our findings in
Section~\ref{sec-summary}.

\section{Simulations}
\label{sec-method}

The simulations presented here are part of the OverWhelmingly Large
Simulations (OWLS) project \citep{Schaye2010}, which consists of a
large suite of cosmological, hydrodynamical simulations that include
chemodynamics. We use a
substantially modified version of the $N$-body 
Tree-PM, smoothed particle hydrodynamics (SPH) code \textsc{gadget
  iii}, which is itself a modified 
version of the code \textsc{gadget ii} \citep{Springel2005}. The code
includes new modules for radiative cooling and heating
\citep{Wiersma2009a}, star formation \citep{Schaye2008}, chemodynamics
\citep{Wiersma2009b}, and feedback from supernovae
\citep{Dallavecchia2008}. Below we will give a brief overview of the
simulation ingredients that are most relevant for this work. More details can be found in \citet{Schaye2010} and the references given above. 

We assume a flat $\Lambda$CDM universe with the cosmological parameters
$[\Omega_m,\Omega_b,\Omega_\Lambda,\sigma_8, n_s, h, Y] = [0.238,
  0.0418, 0.762, 0.74, 0.951, 0.73, 0.248]$, as determined from the
WMAP 3-year data and consistent\footnote{The most significant discrepancy is in
  $\sigma_8$, which is 8\% lower than the value favoured by the WMAP
  7-year data.} with the WMAP 7-year data \citep{Jarosik2010}.

We use the radiative cooling tables presented in 
\cite{Wiersma2009a}\footnote{We used their equation (3) rather than (4)
  and \textsc{cloudy} version 05.07 rather than 07.02.}. In brief, net
radiative cooling rates are computed element-by-element in the
presence of the cosmic microwave background (CMB) and the
\citet{HM01} model for the UV/X-ray background
radiation from quasars and galaxies. The contributions of the eleven
elements hydrogen, helium, carbon, nitrogen, oxygen, neon, magnesium,
silicon, sulphur, calcium, and iron are interpolated as a function of
density, temperature, and redshift from tables that have been
pre-computed using the publicly available photo-ionization package
\textsc{cloudy}, last described by \cite{Feta98}, assuming the gas to
be optically thin and in (photo-)ionization equilibrium.

Stars are formed from gas particles as per \citet{Schaye2008}. 
Gas at sufficiently  
high densities ($n_{\rm H} \ga 10^{-1}\,\cm^{-3}$) is expected to be
multiphase and star-forming \citep{Schaye2004}. Because we lack the
resolution and the physics to model the cold interstellar phase, we 
impose an effective equation of state with pressure $P\propto
\rho_{\rm g}^{4/3}$ for densities exceeding $n_{\rm H} =
0.1~\cm^{-3}$, normalised to $P/k = 1.08\times 10^3~\cm^{-3}\,\K$ at
the threshold. For this slope of the equation of state both the Jeans
mass and the ratio of the Jeans length and the SPH 
kernel are independent of the density, thus preventing spurious
fragmentation due to a lack of numerical
resolution. Star-forming gas particles are stochastically converted into star particles at a pressure-dependent rate that is determined analytically from the observed Kennicutt-Schmidt star formation law.  

Supernova feedback is injected in kinetic form, as described in
\cite{Dallavecchia2008}. After a short delay of $t_{\rm SN} =
3\times 10^7~\yr$, corresponding to the maximum lifetime of stars that
end their lives as core-collapse supernovae, newly-formed star
particles inject kinetic energy into their surroundings by kicking a
fraction of their neighbours in a random direction. The reference simulations
presented here use the default parameters of \cite{Dallavecchia2008},
which means that each SPH neighbour $i$ of a newly-formed star
particle $j$ has a probability of $\eta m_j/\sum_{i=1}^{N_{\rm
    ngb}}m_i$ of receiving a kick with a velocity $v_{\rm w}$, where $N_{\rm
    ngb}=48$ is the number of SPH neighbours. We
choose $\eta = 2$ and $v_{\rm w} = 600~\kms$ (i.e.\ if all baryonic
particles had equal mass, each newly formed star particle would kick,
on average, two of its neighbours). Assuming that each star with
initial mass in the range $6-100~\Msun$ injects $10^{51}~\erg$ of
kinetic energy, these parameters imply that the total wind energy
accounts for 40 per cent of the available kinetic energy for a
Chabrier IMF and a stellar mass range $0.1-100~\Msun$ (if we consider
only stars in the mass range $8-100~\Msun$ for type II supernovae,  this works
out to be 60 per cent). The value $\eta=2$ was chosen to roughly
reproduce the peak in the cosmic star formation rate \citep[see][]{Schaye2010}. 

For the chemo-dynamics, we employ the method described in \citet{Wiersma2009b}.
We use a \cite{C03} stellar initial mass function,
with mass limits of $0.1~\Msun$ and $100~\Msun$. We consider enrichment due to mass loss from AGB and massive
stars as well as type Ia and II supernovae (SNIa and SNII
respectively). Our prescription for stellar evolution is heavily based on
the Padova models as we use the lifetimes of \citet{Portinari1998} and the
yields of low mass and high mass stars of \citet{M01} and \citet{Portinari1998},
respectively. The yields of \citet{Portinari1998} include ejecta from SNII along with estimates of mass loss
from high mass stars. Using these yields gives an element of
consistency between the low and high mass stars. Finally, we use the `W7' SNIa
yields of \cite{Teta03}.

As the precise channels of SNIa progenitors are uncertain, we employ SNIa explosion rates that roughly reproduce the observed cosmic SNIa
rate. We note that the observations from different surveys are not consistent and that our
rates conform to the most recent data.  An e-folding delay time is
used to describe the SNIa rate and we use a
corresponding efficiency (i.e.\ number of white dwarfs that become
SNIa) of $2.5\%$.

We use SPH-smoothed
metallicities for cooling and stellar evolution purposes, where $Z_{\rm sm} \equiv \rho_{\rm Z}/\rho_{\rm g}$ for gas particles, and star particles inherit the smoothed metallicities of their progenitors. In \citet{Wiersma2009b} we argued that this
definition is more consistent with the SPH formalism than the more commonly used particle metallicity, $Z_{\rm part} \equiv m_{\rm Z}/m_{\rm g}$. Using smoothed
metallicities results in a spreading of metals over greater volumes, increasing the 
average rates of cooling and star formation.

Since SPH is a Lagrangian scheme, metals move with the particles they
are attached to, and hence a high-metallicity particle may in principle
be completely surrounded by low-metallicity gas. Introducing smoothed
metallicities partially mitigates this numerical artefact. However,
SPH codes may underestimate the true amount of gas mixing as compared to grid codes in case of instabilities  \citep{Agertz2007} or turbulence
\citep{Mitchell2009}. It would thus be very interesting to investigate whether the metal distribution in such grid simulations is similar to that obtained with SPH. On the other hand, we note that \cite{Shen2009} included turbulent diffusion of metals in cosmological SPH simulations but found its effect to be small.

\begin{table*}
\caption{List of simulations. From left-to-right the column show: simulation identifier, dark matter particle mass, baryonic particle mass, maximum physical gravitational softening (the softening is held fixed in comoving coordinates above $z=2.91$), final redshift, and a brief description.}  
\centering
\label{tab-simset}
\begin{tabular}{lcccll}
\hline\hline
Simulation & $m_{\rm dm}$ & $m_{{\rm b}}$ & $\epsilon_{\rm prop}$ & $z_{{\rm end}}$ & Comment  \\ 
& $(\hMsun)$ & $(\hMsun)$ & $(\hkpc)$ \\
\hline
\textit{L025N512} & $6.3 \times 10^6$ & $1.4 \times 10^6$ & 0.5 & 1.45 & Reference model \\
\textit{L050N512} & $5.1 \times 10^7$ & $1.1 \times 10^7$ & 1.0 & 0 & Reference model \\
\textit{L025N256} & $5.1 \times 10^7$ & $1.1 \times 10^7$ & 1.0 & 2 & Reference model \\
\textit{L050N256} & $4.1 \times 10^8$ & $8.7 \times 10^7$ & 2.0 & 0 & Reference model \\
\textit{L100N512} & $4.1 \times 10^8$ & $8.7 \times 10^7$ & 2.0 & 0 & Reference model \\
\textit{AGN\_L100N512} & $4.1 \times 10^8$ & $8.7 \times 10^7$ & 2.0 & 0 & Includes AGN \\
\textit{WDENS\_L100N512} & $4.1 \times 10^8$ & $8.7 \times 10^7$ & 2.0 & 0 & Wind vel.\ prop.\ to local sound speed (SN energy as \emph{REF})\\
\hline
\end{tabular}
\end{table*}

Of the OWLS suite of simulations \cite[described in][]{Schaye2010},
we use the subset listed in Table~\ref{tab-simset}. Here we make use
of the notation \textit{LXXXNYYY}, where the simulation cube has a comoving
side length of XXX~$\hMpc$ and uses YYY$^3$ dark matter particles and
YYY$^3$ baryonic particles. Table~\ref{tab-simset} also lists the dark matter and the initial baryonic
particle masses, $m_{\rm dm}$ and $m_{\rm b}$, the maximum proper
gravitational softening (the softening is fixed to this value in comoving
coordinates above $z=2.91$), $\epsilon_{\rm prop}$, the redshift at which the simulation was
stopped, $z_{\rm end}$, and a short description. 

We explore two variations on the reference model (\textit{REF}). Run
\textit{AGN\_L100N512} includes the prescription of \citet{Booth2009}
for the growth of supermassive black holes by gas accretion and
mergers, and for feedback from AGN. The black holes inject 1.5~per
cent of the accreted rest mass energy as thermal energy in the
surrounding interstellar medium. This model reproduces not only the
the local black hole scaling relations \citep{Booth2009},
but also the observed entropy, temperature, and metallicity profiles
of groups of galaxies and the stellar masses that they contain
\citep{McCarthy2009}.  

For run
\textit{WDENS\_L100N512} the parameters of the winds are dependent
on the local density, keeping the energy per unit stellar mass formed constant.  They scale such that in high-density
environments the wind velocity is higher:
  \begin{eqnarray}
    v_{\rm w} &=& v_{\rm w}^\ast \left (\frac{n_{\rm H}}{n_{\rm H}^\ast} \right )^{1/6}, \\ 
\eta &=& \eta_\ast \left (\frac{n_{\rm H}}{n_{\rm H}^\ast} \right )^{-1/3},
  \end{eqnarray}
where $v_{\rm w}^\ast=600~\kms$ and $\eta_\ast = 2$ such that the model agrees with the reference model at the threshold for star formation. For higher gas pressures the velocity increases in proportion with the local sound speed. \cite{Schaye2010} find that this model is much more efficient at suppressing star formation in massive galaxies, though not as efficient as the model that includes AGN.

\begin{figure}
\includegraphics[width=84mm]{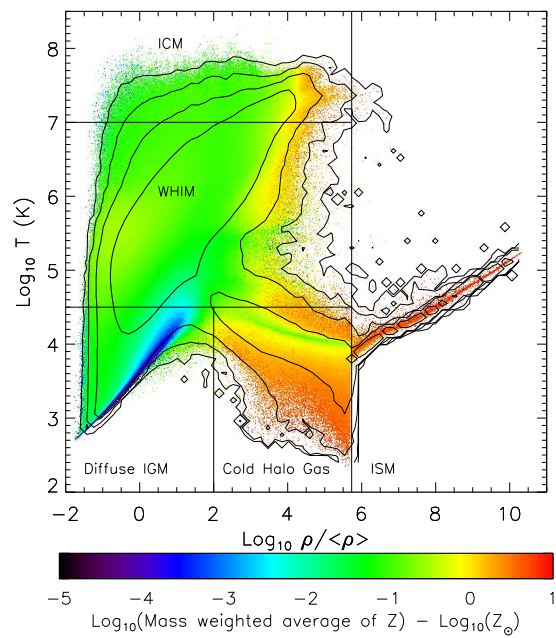}
\caption{Mass weighted metal distribution in temperature-density space at $z =
  0$ in simulation \textit{L050N512}. The colour scale gives the metallicity. The contours indicate the metal mass distribution and are
  logarithmically spaced by 1 dex. The straight lines indicate the adopted division of the gas into: non-star-forming gas (i.e., $n_{\rm
H}<0.1~\cm^{-3}$), diffuse photo-ionised IGM ($\rho < 10^2\,\langle \rho \rangle$, $T<10^{4.5}\,\K$),
  cold halo gas ($\rho>10^2\,\langle \rho \rangle$,
$T<10^{4.5}\,\K $), WHIM ($10^{4.5}\,\K < T <
10^7\,\K$), and ICM ($T > 10^7\,\K$). The metals are
  distributed over a wide range of densities and temperatures. \label{fig-smZcont}}
\end{figure}

\section{When was the gas enriched?}
\label{sec-when}

In this section we will
investigate how the time at which a gas element was enriched varies with
its density and temperature at some later time. We will evaluate the
enrichment history of the gas as a function of its physical state at
both $z=0$ and $z=2$. Before doing so, it is, however, instructive to look
at the distribution of metal mass at the present time, which we show
in Fig.~\ref{fig-smZcont}. 

Fig.~\ref{fig-smZcont} indicates how we can divide the
temperature-density plane in several regions of interest. For $n_{\rm
  H}>0.1~\cm^{-3}$ (which corresponds to $\rho/\langle\rho\rangle >
5.3 \times 10^5$ at $z=0$ and $\rho/\langle\rho\rangle>2.0\times 10^4$
at $z=2$) we impose an effective equation of state and assume the gas
to be star-forming. We divide the non-star-forming gas into hot gas
typically found in haloes of large groups or clusters (ICM, $T >
10^7\,\K$), warm-hot intergalactic gas (WHIM, $10^{4.5}\,\K < T <
10^7\,\K$), diffuse photo-ionised IGM ($\rho < 10^2\,\langle \rho
\rangle$, $T< 10^{4.5}\,\K$), and cold halo gas ($\rho>10^2\,\langle
\rho \rangle$, $T<10^{4.5}\,\K $). While we have divided the cold ($T
< 10^{4.5}\,\K$) gas into low- and high-density components (diffuse
IGM and cold halo gas, respectively), we have not done the same for
hotter gas. We find that if we split the WHIM at $\rho = 10^2\,
\langle \rho \rangle$, the enrichment history of the low-density
component is very similar to that of the WHIM in total, whereas that of
the high-density component is more similar to the enrichment history
of the ICM. The high-density component contains less than one quarter of the WHIM metals at $z = 0$ and less than one sixteenth at $z=2$.

\begin{figure*} 
\includegraphics[width=84mm]{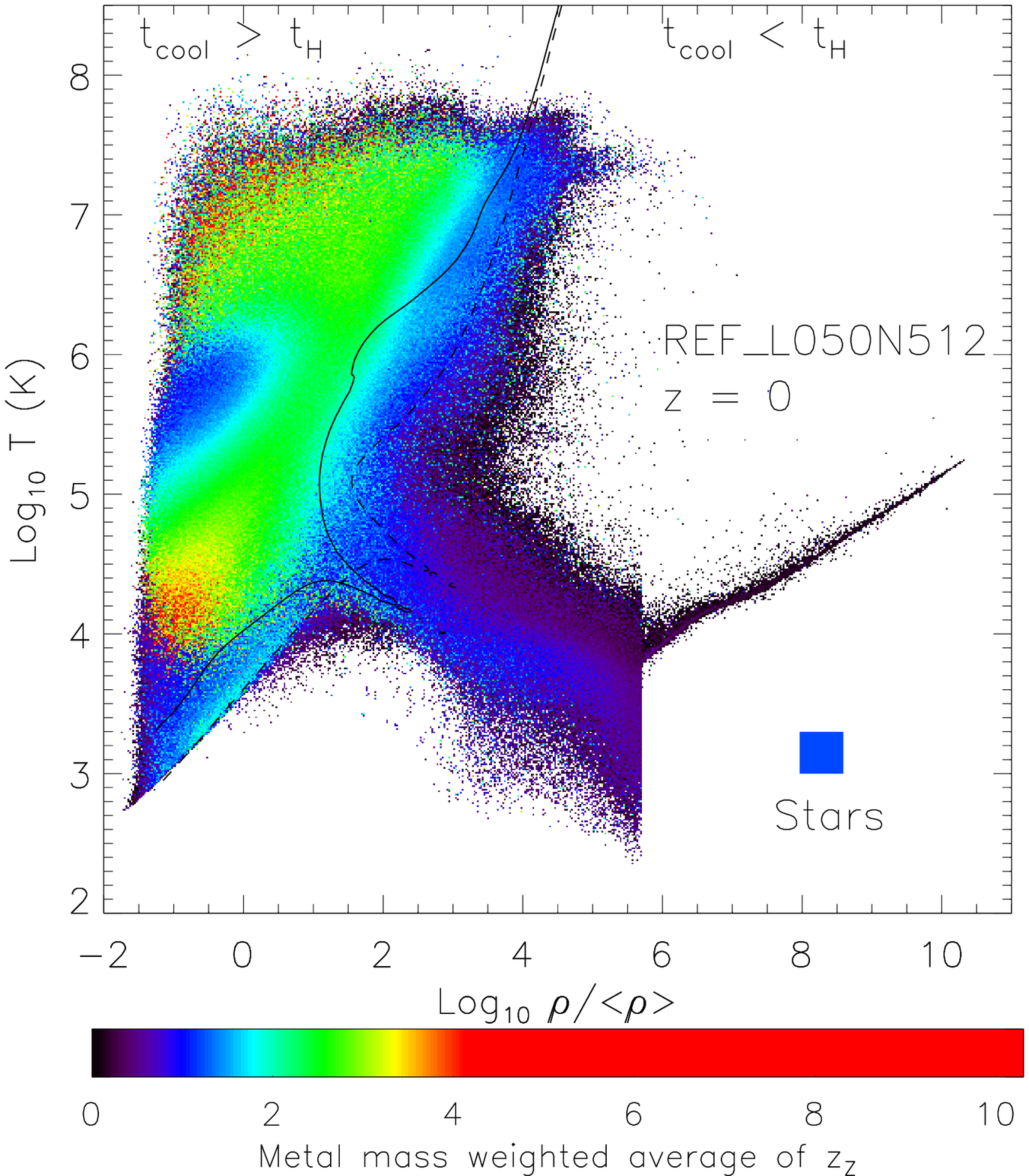}
\includegraphics[width=84mm]{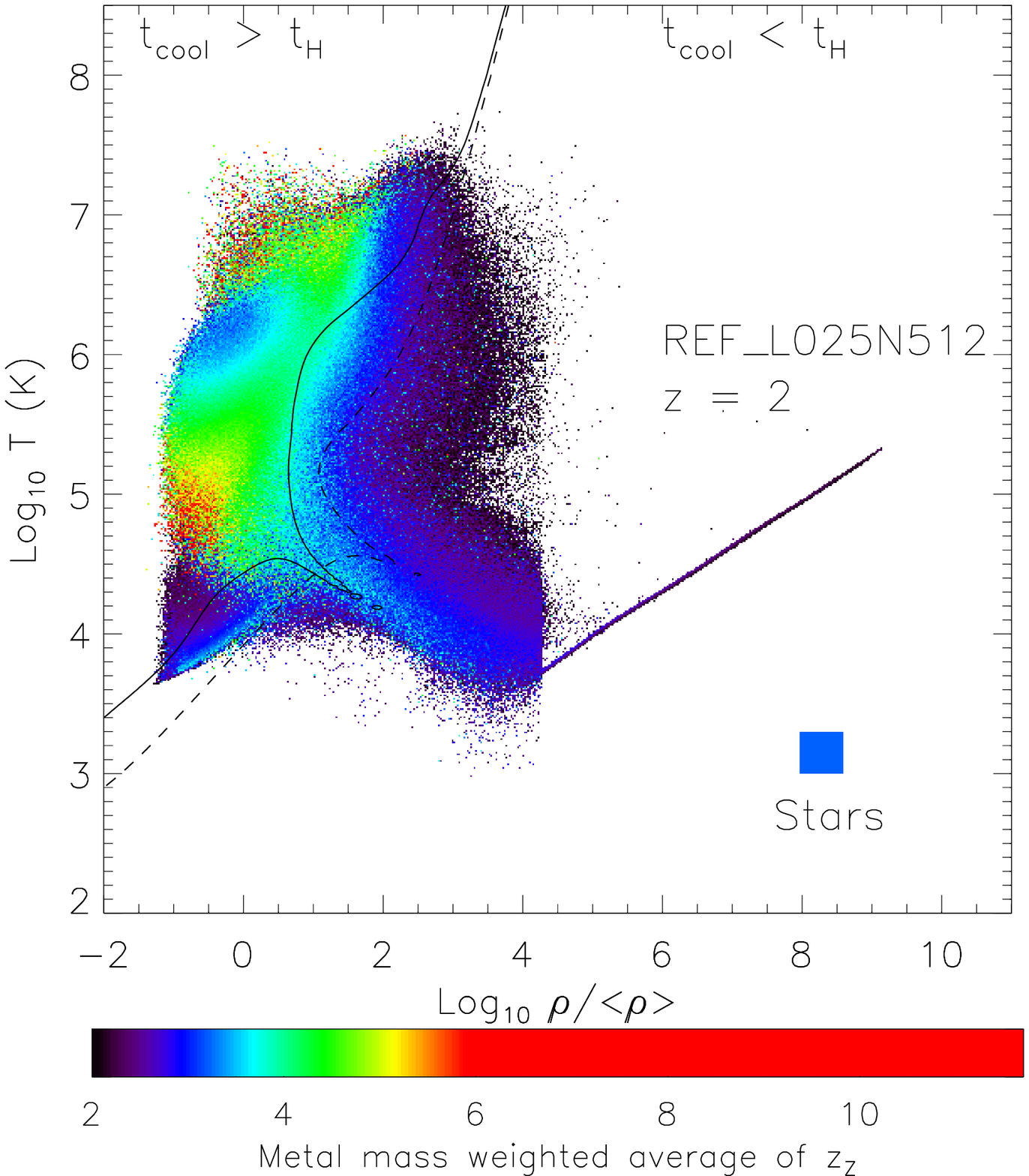}
\caption{Enrichment redshift in temperature-density space. The colour scale gives the metal mass weighted mean enrichment redshift as a function of the gas density and temperature at $z = 0$ in \textit{L050N512} (left panel) and at $z=2$ in \textit{L025N512} (right panel). Each pixel shows the
  metal mass weighted average value of $z_Z$ for all particles that fall into the corresponding temperature-density bin. The rectangles in the bottom-right parts of the diagrams indicate
the corresponding values for the metals locked up in stars. Also
  plotted are $t_{\rm cool} = t_{\rm H}$ contours for solar (solid) and primordial 
  (dashed) abundances. Note that we have stretched the colour scale to emphasise the different components (only a handful of pixels correspond to high values of $z_Z$) and that the two panels use different colour scales. Denser gas was typically enriched later. The enrichment redshift is particularly small for gas with a short cooling time.  \label{fig-L050zwz}}
\end{figure*}

In \citet{Wiersma2009b} we showed that the metallicities predicted for
these phases\footnote{Note that in \citet{Wiersma2009b} we used
  $T=10^5\,\K$ as the dividing temperature for WHIM and cold gas; we
  lower the limit here so that we can better isolate the components
  visible in Fig.~\ref{fig-L050zwz}.} by our reference model are in
agreement with the observations, although the metallicity for
the diffuse IGM may be slightly too low and is not yet fully converged
with respect to the numerical resolution (it increases with increasing
resolution).

The contours in Fig.~\ref{fig-smZcont} show how the metal mass is distributed in temperature-density space. The metals are spread over a wide range of phases. The metallicity (that is, the ratio of metal mass to total mass), indicated by the colour scale, ranges significantly between the various regions. That the contours do not always trace the regions with the highest metallicities highlights the fact that the metals are distributed rather differently from the rest of the baryons. 
Table \ref{tab-metfrac} summarises the $z = 0$ and $z = 2$ metal mass distributions for the reference model. 

\begin{table}
\caption{Fractional distribution of metals in the reference models at $z=0$ and $z=2$.}  
\centering
\label{tab-metfrac}
\begin{tabular}{lr}
\hline\hline
Phase & Fraction of metals (\%)\\
\hline
\multicolumn{2}{c}{L050N512 ($z = 0$)}\\
\hline
Stars & 68.2\\
SF Gas & 5.8\\
ICM & 1.3 \\
WHIM & 19.0 \\
Cold Halo Gas & 4.6 \\
Diffuse IGM & 1.0 \\
\hline
\multicolumn{2}{c}{L025N512 ($z = 2$)}\\
\hline
Stars & 42.0\\
SF Gas & 25.8\\
ICM & $<$ 0.1 \\
WHIM & 24.0 \\
Cold Halo Gas & 7.2 \\
Diffuse IGM & 0.9 \\
\hline
\end{tabular}
\end{table}

We will make frequent use of what we will term ``metal-mass-weighted mean redshifts'', $z_Z$. This quantity, which 
we compute on the fly for all gas particles in our simulations, is defined as the mean redshift of enrichment events weighted by the metal mass imparted by each event:
\begin{equation}
z_Z \equiv \frac{\sum_i \Delta m_{{\rm Z};i} z_i}{m_{\rm Z}},
\end{equation}
where $\Delta m_{Z;i}$ is the metal mass received by the gas particle at
redshift $z_i$ ($i$ indicates a time-step) and $m_Z$ is the total metal mass of the 
particle. Thus, if a gas particle received all its metals over 
a short interval (which is typical for particles that have been ejected from haloes), this quantity gives the redshift at which   
it was enriched. Note that $z_Z$ does not correspond to the time at which the metals were created. Stars may create metals long before they are released and metals may be cycled through multiple generations of stars. We will nevertheless sometimes refer to the metals residing in gas with higher values of $z_Z$ as being ``older''. In \S\ref{sec-what} we will show that for intergalactic metals $z_Z$ is typically close to the redshift at which the metals last resided in a halo. 

Fig.~\ref{fig-L050zwz} shows $z_Z$ in temperature-density space for \textit{L050N512} at
$z = 0$ (left panel) and 
for \textit{L025N512} at $z = 2$ (right panel). 
Note that this figure contains no information about how much metal mass is
actually in a given bin (for instance, the diffuse IGM contains
relatively little metal mass, but this is not obvious from this
plot). We can, however, refer back to 
Fig.~\ref{fig-smZcont} to see the metal distribution. For reference,
computing the metal mass weighted average over all gas (stars) yields $z_Z\approx 1.5$ ($z_Z\approx 1.2$) for \textit{L050N512} at $z=0$ and $z_Z\approx 3.1$ ($z_Z\approx 3.2$) for
\textit{L025N512} at $z=2$. 

The most prominent trend visible in Fig.~\ref{fig-L050zwz} is a strong overall gradient with density, with higher density gas being enriched more recently. This agrees with the predictions of \citet{Oppenheimer2009} for the low-redshift, diffuse IGM as traced by \OVI\ absorbers.
Of all components, the star-forming gas (i.e., the ISM) has been enriched most recently. This is not unexpected, as this is the gas that surrounds the stars that are releasing the metals. Moreover, much of the metal mass that was injected into the ISM a long time ago, will already have been locked up in stars. 
Indeed, the metals locked up in stars were, on average, released (by other, older stars) earlier than the metals that reside in the ISM as can be seen by comparing the colour of the rectangles in the bottom-right parts of the panels to that of the ISM. 
The fact that the increase of $z_Z$ with decreasing density continues all the way to the lowest densities, suggests that travel-time is a limiting factor for the enrichment of the IGM, as suggested by \citet{Aguirre2001z3}. 

Fig.~\ref{fig-L050zwz} also reveals some interesting trends with the gas temperature. For high gas densities, the dependence on temperature reflects a dependence on the radiative cooling time. We can see this by comparing to the solid and dashed contours, which show where the net radiative cooling time\footnote{We plot only the radiative cooling time, but note that the actual cooling rates in the simulation are dominated by adiabatic expansion for low densities.}, $t_{\rm cool}$, equals the Hubble time, $t_{\rm H}$, for gas with solar and primordial abundances, respectively. The two contours that enter the diagram at high and low temperatures correspond to net cooling and heating, respectively. They 
converge at high densities and intermediate temperatures at the equilibrium temperature. There the contours should really extend further to the bottom right, approximately to the star formation threshold (above which we impose an effective equation of state), but they stop due to the finite resolution of the grid on which we have computed the cooling time. Gas to the right of the contours has $t_{\rm cool} < t_{\rm H}$, but gas to the left of the contours cannot cool. Clearly, the enrichment redshift is a decreasing function of the cooling time if $t_{\rm cool} < t_{\rm H}$. Since the gas cannot remain in the same area of the temperature-density plane for more than a cooling time, this implies that the time since enrichment is correlated with the time since the gas parcel moved significantly in the $T-\rho$ diagram. This is for example expected if the gas was enriched and shock-heated at roughly the same time, as in a galactic wind. It would, however, also happen if the enrichment led to a strong reduction in the cooling time.

\begin{figure*}
\includegraphics[width=168mm]{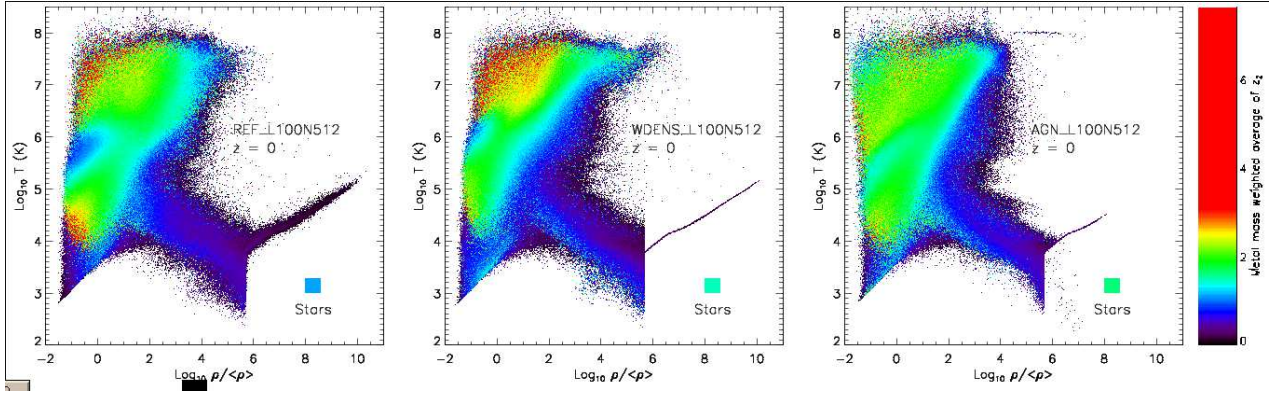}
\caption{A comparison of the enrichment redshifts for simulations including different feedback processes. The colour scale gives the metal mass weighted mean enrichment redshift as a function of the gas density and temperature at $z=0$ for simulations \textit{REF\_L100N512} (left panel), \textit{WDENS\_L100N512} (middle panel), 
  and \textit{AGN\_L100N512} (right panel). Model \textit{WDENS} uses an implementation of supernova-driven winds that is more efficient in massive haloes than that of model \textit{REF}. Model \textit{AGN} uses the same prescription for supernova feedback as model \textit{REF}, but includes feedback from AGN. 
The rectangles in the bottom-right parts of the diagrams indicate the values of $z_Z$ for the metals locked up in stars. For numerical reasons that are explained in the main text, the unresolved multiphase ISM (Log $\rho/\langle\rho\rangle \ga 6$) follows a much tighter equation of state for models \textit{WDENS} and \textit{AGN} than for model \textit{REF}. While the features seen at $(\rho,T)\sim (10^{-1}\,\left < \rho \right >,10^{4.3}\,\K)$ and 
$(10^{-0.5}\,\left < \rho \right >,10^{5.7}\,\K)$ for model \textit{REF} depend on the implementation of galactic winds, the other trends are robust. \label{fig-L100zwz}}
\end{figure*}

In the long cooling time regime (i.e.\ to the left of the contours) the enrichment redshift also depends on the gas temperature. We will show later (see Fig.~\ref{fig-L100zwz}) that some of the features for $\rho \la \left < \rho \right >$ and $T > 10^4\,\K$ are sensitive to the implementation of feedback. However, the finding that the coldest part of the diffuse IGM was enriched more recently than the WHIM, is robust. Note, however, that the diffuse IGM contains very few metals compared to the WHIM (see Fig.~\ref{fig-smZcont}). We have checked and have 
found that most of the metals in the cold, underdense part of the diffuse IGM have never been part of a (resolved) halo. This suggests that this gas may have been enriched by winds from intermediate-mass, intergalactic stars (presumably formed in galaxies and then ejected). Apparently, gas that has been enriched and ejected by galactic winds ends up in the WHIM at higher temperatures, probably because it has been put onto a higher adiabat as a result of shock-heating. Once the gas has entered the region where $t_{\rm cool} > t_{\rm H}$, it can no longer decrease its entropy, preventing low-density WHIM gas from moving into the diffuse, photo-ionised region of the diagram. 

The left panel of Fig.~\ref{fig-L100zwz} shows $z_Z(T,\rho)$ at $z=0$ for the \textit{L100N512} version of the reference model. Comparison with the left panel of Fig.~\ref{fig-L050zwz}, which showed the prediction for a simulation with 8 (2) times higher mass (spatial) resolution shows that the trends have converged numerically.
 
The remaining panels of Fig.~\ref{fig-L100zwz} show the enrichment redshift predicted by two other simulations from the OWLS project. The middle panel shows model \textit{WDENS}, for which the wind
velocity scales with the local sound speed, while keeping the total energy per unit stellar mass identical to that used in model \textit{REF}. In this model the supernova driven winds remain efficient up to higher halo masses, which results in a lower global star formation rate at low redshift \citep{Schaye2010}. The right panel shows the results for model \textit{AGN}, which uses the same prescription for supernova feedback as model \textit{REF}, but which also includes feedback from accreting supermassive black holes. Since AGN feedback is most relevant for high halo masses, the star formation rate, and thus the rate at which metals are both produced and locked up in stars, decreases much faster below $z=2$ than for the other models \citep{Schaye2010}. This explains the increase in the metal mass weighted mean enrichment redshift of the stars, from $z_Z \approx 1.1$ for model \textit{REF} to 1.5 for \textit{WDENS} to 1.6 for \textit{AGN} (the corresponding values for the gas are $z_Z \approx 1.1$, 1.3, and 1.5, respectively).

The gas on the equation of state (Log $\rho/\langle\rho\rangle \ga 6$), which represents the unresolved, multiphase ISM, displays much less scatter for \textit{WDENS} and \textit{AGN} than for \textit{REF}. This reflects the fact that for the latter model the output time corresponded to a global time-step whereas passive particles were caught `drifting' adiabatically between time-steps in the reference run, resulting in temperatures that deviate slightly from our imposed equation of state. Note that this does not represent any difference in how the
hydrodynamical forces were calculated among the three
simulations. Observe also that some of the particles in the right
panel fall on a line near $10^8\,\K$. This corresponds to the AGN heating temperature \citep[see][]{Booth2009} and thus implies that these particles have been heated very recently.

Comparing the three panels, we see that the features seen at $(\rho,T)\sim (10^{-1}\,\left < \rho \right >,10^{4.3}\,\K)$ and 
$(10^{-0.5}\,\left < \rho \right >,10^{5.7}\,\K)$ for model \textit{REF} depend on the implementation of supernova-driven winds. All other trends of $z_Z$ with $T$ and $\rho$ are, however, robust. Even AGN feedback does not change the trends discussed above. 
Since the differences between the simulations are small compared to the similarities, we will only make use of our higher resolution reference model for the remainder of this section. It should be kept in mind, however, that this model probably underestimates the typical enrichment redshift due to the absence of an efficient feedback mechanism in massive galaxies. 

\begin{figure*} 
\resizebox{\colwidth}{!}{\includegraphics{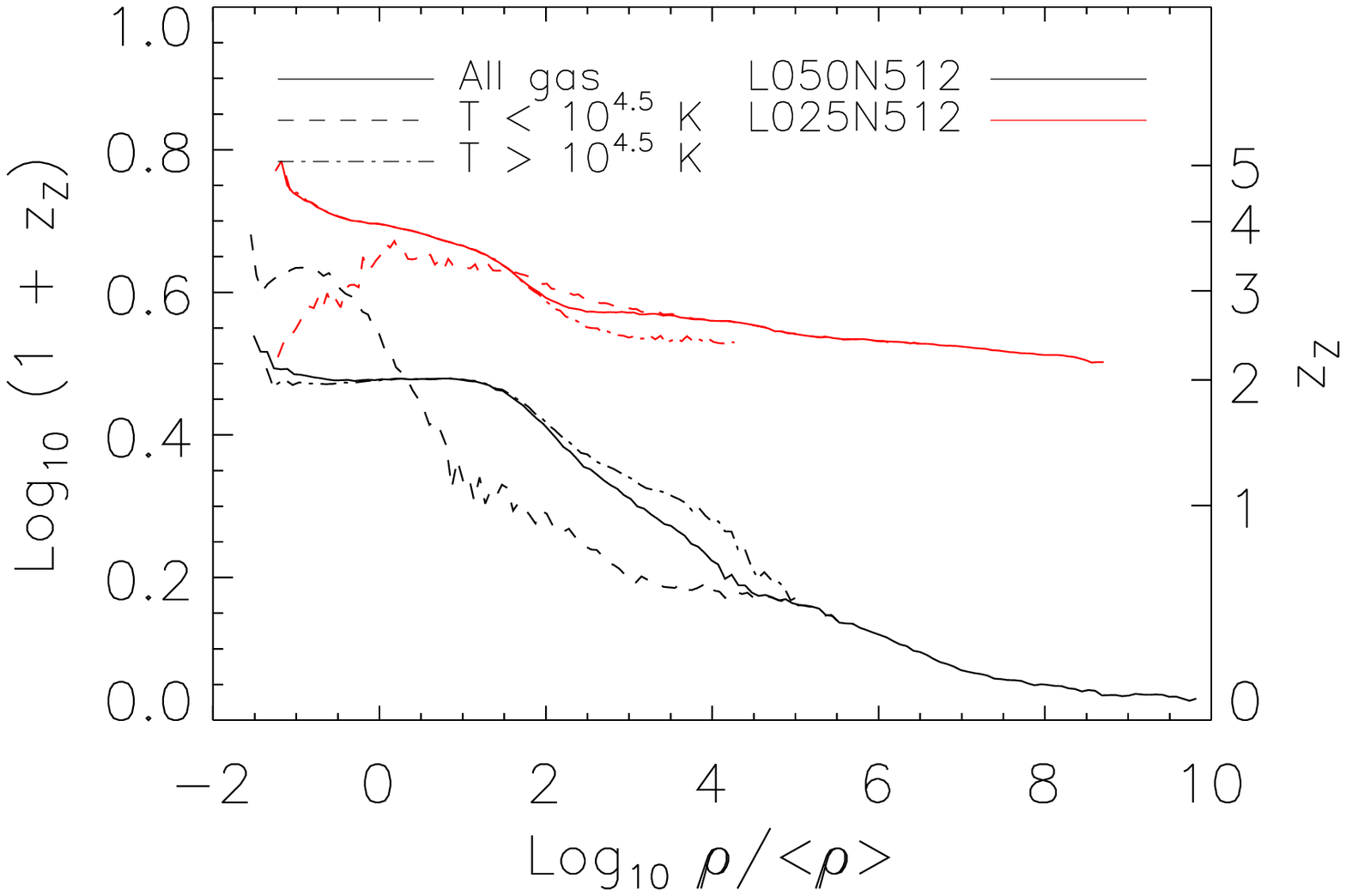}}
\resizebox{\colwidth}{!}{\includegraphics{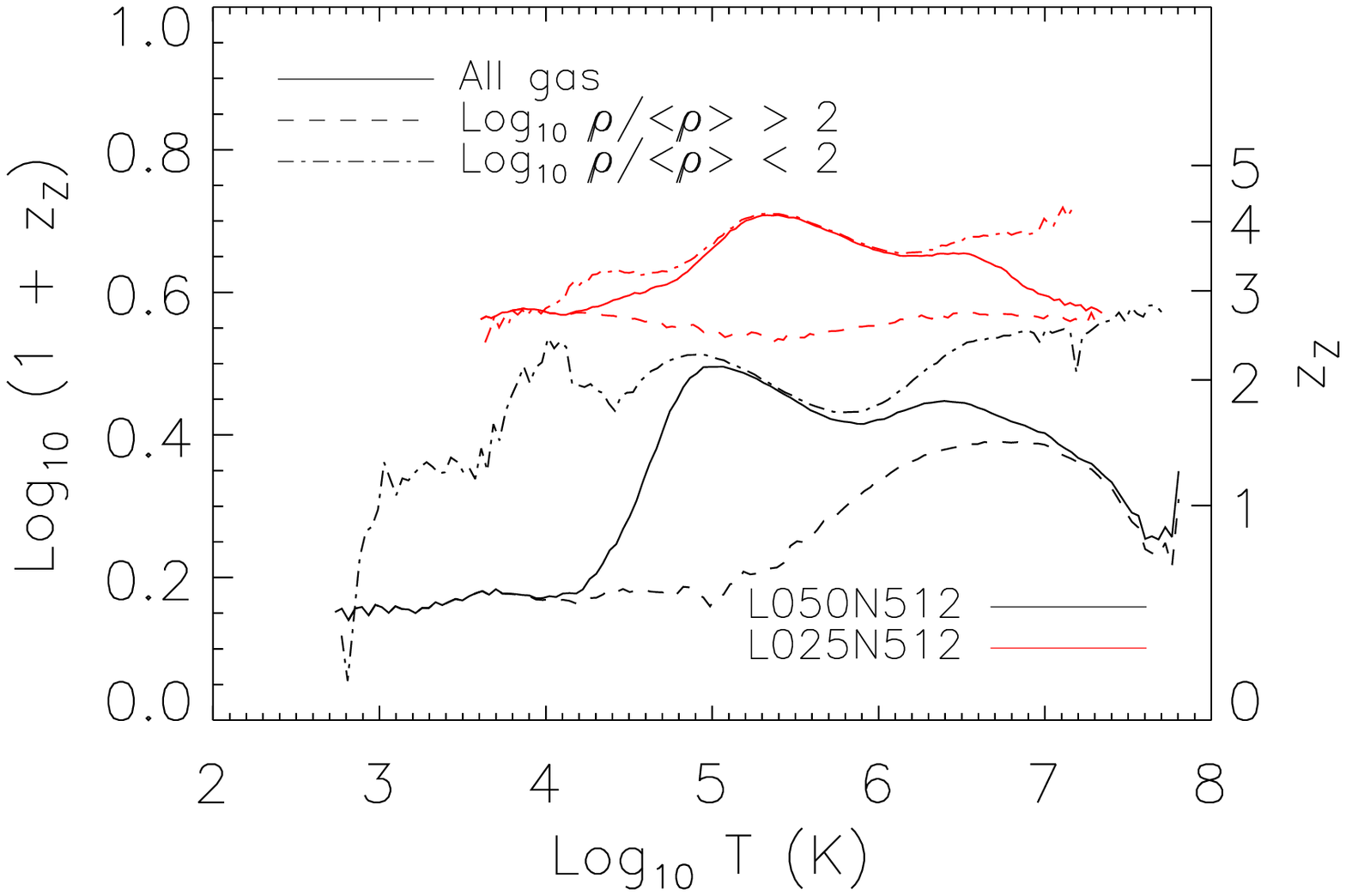}}
\caption{Metal mass weighted mean enrichment redshift as a function of the gas density (left panel) or the gas temperature (right panel) at $z=0$ for model \textit{L050N512} (black) and at $z=2$ for \textit{L025N512} (red). The different line styles correspond to cuts in the temperature (left panel) or density (right panel) as indicated in the legends. The curves were cut when the
number of particles per bin falls below 200. The enrichment redshift decreases rapidly with increasing density. Metals residing in gas with intermediate temperatures ($T\sim 10^5 - 10^6\,\K$) are typically older, but this merely reflects the fact that metals at these temperatures reside mostly in lower density gas than metals at lower or higher temperatures, as can be seen by comparing the different line styles in the right panel.  \label{fig-zwzcontour}}
\end{figure*}

Figs.~\ref{fig-L050zwz} and \ref{fig-L100zwz} do not allow us to infer what the typical enrichment time is of metals as a function of just density or just temperature, because the diagrams do not take into account the distribution of metal mass in the $T-\rho$ plane.  The solid curves in Fig.~\ref{fig-zwzcontour} therefore show the metal mass weighted mean enrichment redshift as a function of density (left panel) or temperature (right panel). The black and red curves show results for \textit{L050N512} at $z=0$ and for \textit{L025N512} at $z=2$, respectively. 

Focusing first on the left panel, we see strong gradients with the gas density. While gas that is diffuse at $z=0$ was typically enriched at $z\approx 2$, metals that reside in collapsed structures (i.e.\ $\rho \ga 10^2 \left < \rho \right >$) at the present time, were on average released much more recently. The enrichment redshift continues to decrease down to the highest densities. The same holds at $z=2$, although the gradient is less steep at high densities. Gas that is diffuse at $z=2$ typically received its metals at $z>3$ and metals residing in  underdense gas at $z=2$ were on average ejected at $z>4$. 

Similarly, the right panel of Fig.~\ref{fig-zwzcontour} shows the metal mass weighted enrichment redshift as a function of the gas temperature at $z=0$ (black) and $z=2$ (red) for all gas (solid) as well as for gas with density smaller (dot-dashed) or greater (dashed) than $10^2\,\left <\rho \right >$. The enrichment redshift peaks for $T\sim 10^5 - 10^6\,\K$, but this merely reflects the more fundamental trend with density: metals in this temperature range mostly reside in diffuse gas while lower and higher temperatures are dominated by higher densities.

\begin{figure*} 
\includegraphics[width=168mm]{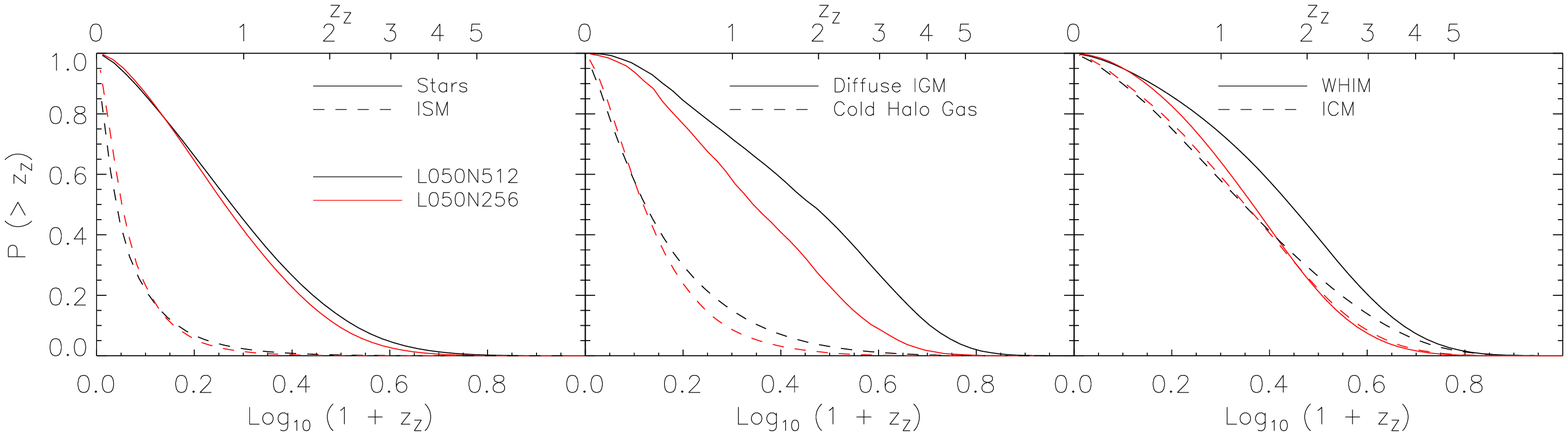}\\ 
\includegraphics[width=168mm]{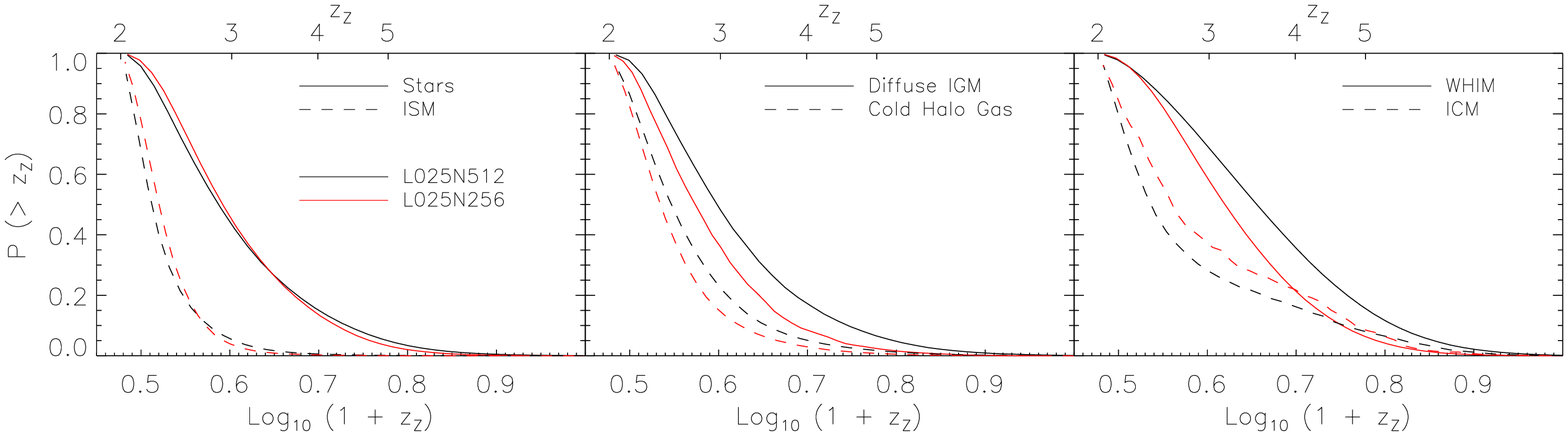}
\caption{Cumulative probability distribution of the enrichment redshift, weighted by metal mass, of metals at $z=0$ (top row) and $z=2$ (bottom row) for stars and for the different gas phases defined in Fig.~\protect\ref{fig-smZcont}. The different colours correspond to different numerical resolutions. The results are converged with respect to resolution except for the diffuse IGM and the WHIM. For this gas even our highest resolution runs may underestimate the enrichment redshift. There is substantial scatter in the enrichment redshift, particularly for the phases whose metals are, on average, older. \label{fig-enrichhist}}
\end{figure*}

\begin{figure*} 
\includegraphics[width=84mm]{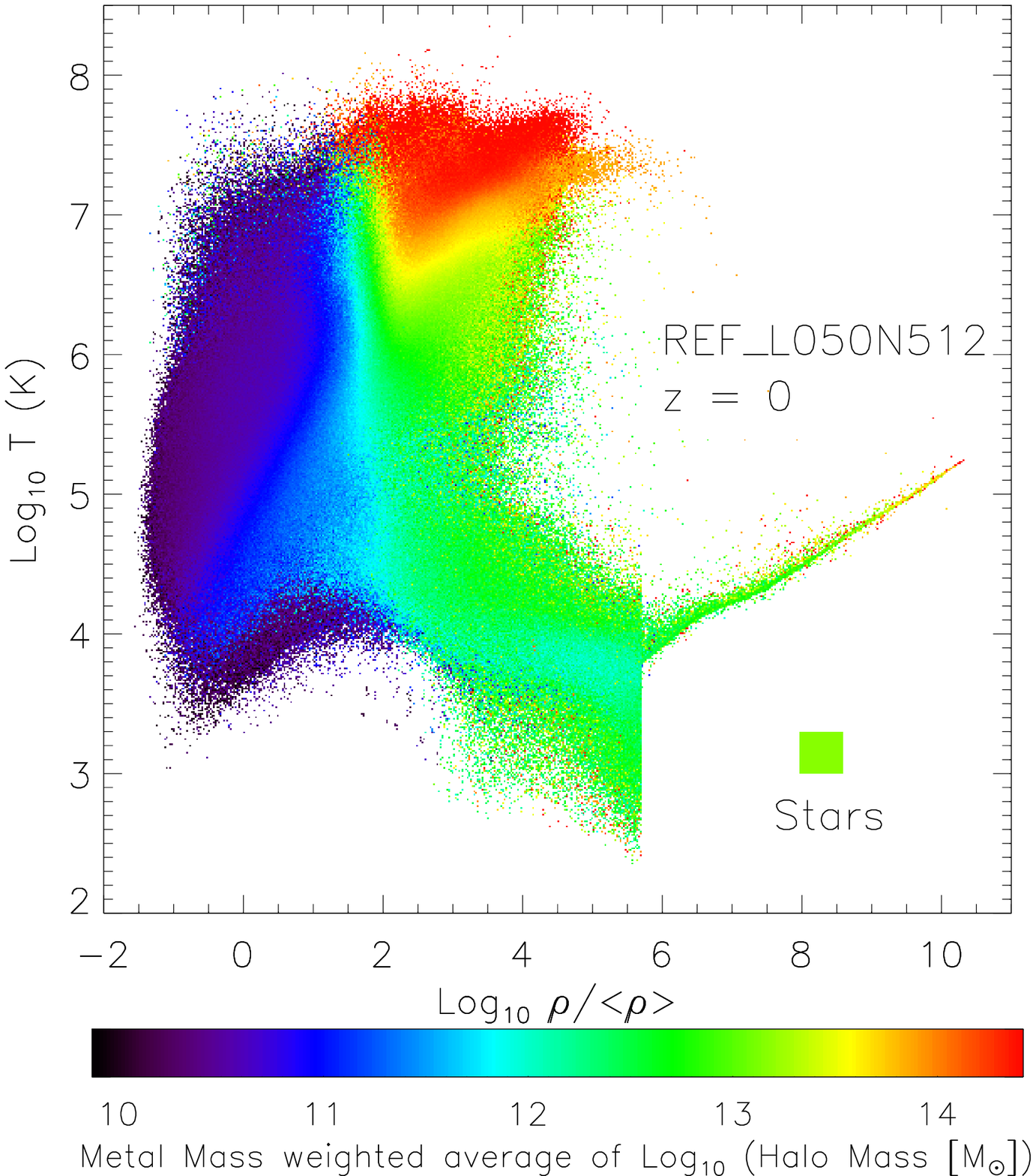}
\includegraphics[width=84mm]{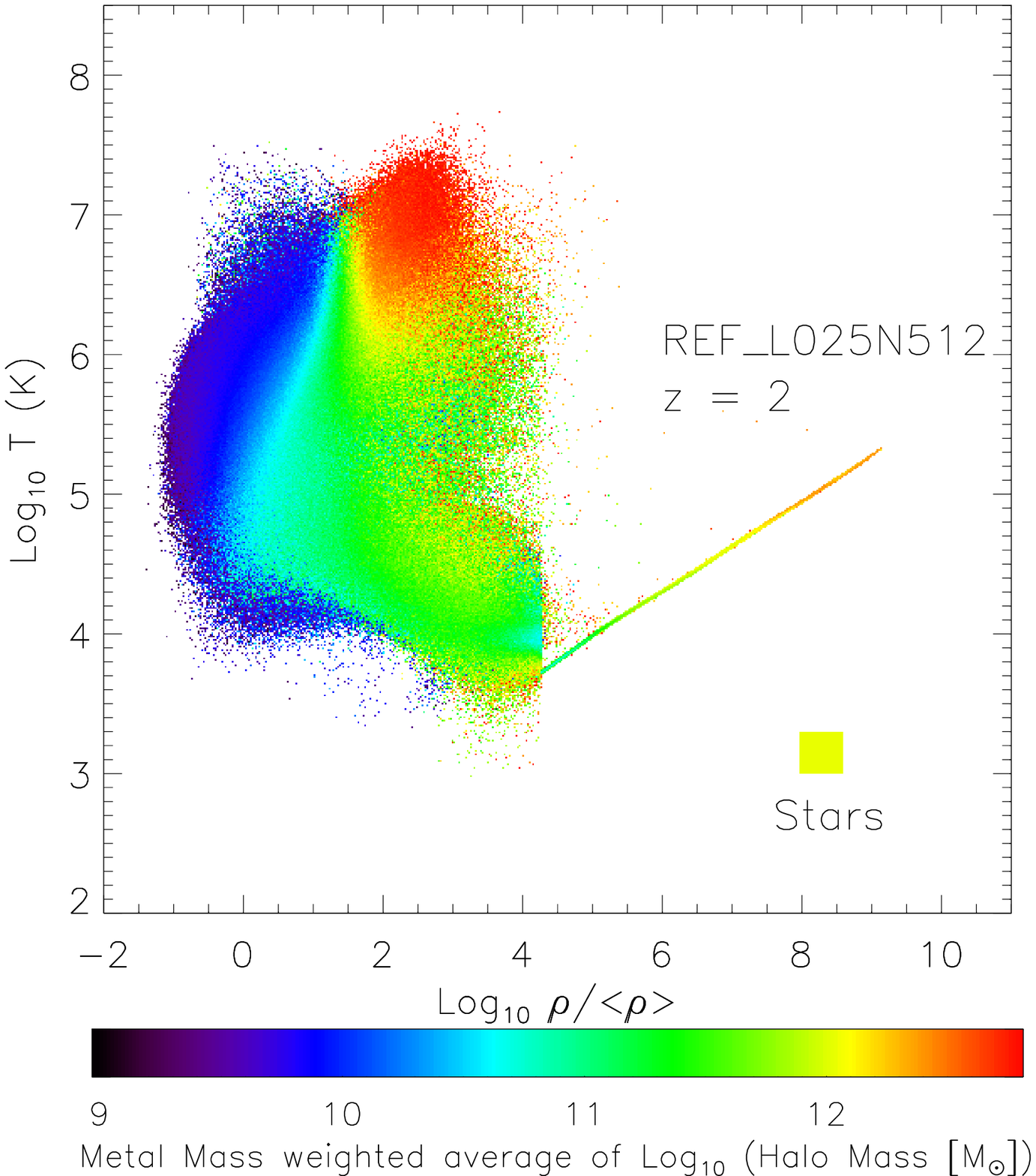}
\caption{Metal mass weighted mean $\log_{10} m_{\rm halo}$, where $m_{\rm halo}$ is the total mass of the halo that last contained the metals, as a function of the gas density and temperature at $z=0$ in \textit{L050N512} (left panel) and at $z=2$ in \textit{L025N512}. Each pixel shows the
  metal mass weighted average value of $\log_{10} m_{\rm halo}$ for all particles that fall into the corresponding temperature-density bin. The rectangles in the bottom-right parts of the diagrams indicate
the corresponding values for the metals locked up in stars. Only particles that have resided in haloes with more than 100 dark matter particles are included. Note that the two panels use different colour scales. Metals residing in lower density gas originate from lower mass haloes. Because high-density gas ($\rho \ga 10^2  \left < \rho \right >$) currently resides in haloes, the increase in halo mass with temperature reflects their virial temperatures. The diffuse IGM is nearly completely missing from this plot because it is enriched by unresolved haloes and/or intergalactic stars.  \label{fig-hmass}}
\end{figure*}

So far we have only plotted the metal mass weighted enrichment redshift averaged over all gas particles in some temperature and/or density bin. Such plots do not reveal information about the distribution in the values of $z_Z$ of the particles that fall into the bin. Fig.~\ref{fig-enrichhist} provides this information by plotting the metal mass weighted cumulative probability distribution function of $z_Z$. In other words, it shows the fraction of the metal mass that resides in gas particles for which $z_Z$ exceeds the value that is plotted along the $x$-axis. The different columns and line styles show the different regions in $T-\rho$ space defined in Fig.~\ref{fig-smZcont} and the different colours correspond to different resolutions. The top and bottom rows are for $z=0$ and 2, respectively. 

Fig.~\ref{fig-enrichhist} shows that the metals residing in a particular gas phase typically span a wide range of enrichment redshifts. Note that this plot underestimates the true range, because it shows the distribution of $z_Z$ for gas particles, which have themselves been averaged over all past enrichment events (weighted by metal mass). That is, even though a particle may have experiences multiple enrichment events, it only has a single value of $z_Z$.
The distribution is particularly wide if the average enrichment redshift is high. While the average value of $z_Z$ is about 2 for the diffuse IGM at $z=0$, a quarter of the metal mass resides in particles with $z_Z< 1$ and another quarter has $z_Z>3$. 

Comparison of the different line colours shows that the convergence with numerical resolution is excellent for stars, the ISM, cold halo gas, and the ICM, but that it is less good for the diffuse IGM and the WHIM, particularly at $z=0$. It is likely that our highest resolution simulation at this redshift, \textit{L050N512}, still significantly underestimates the typical enrichment redshift of these low-density components. We will show in the next section that this is because low-mass haloes dominate the enrichment of low-density gas. 

\section{How massive were the objects that enriched the gas?}
\label{sec-what}

\begin{figure*} 
\includegraphics[width=52.5mm]{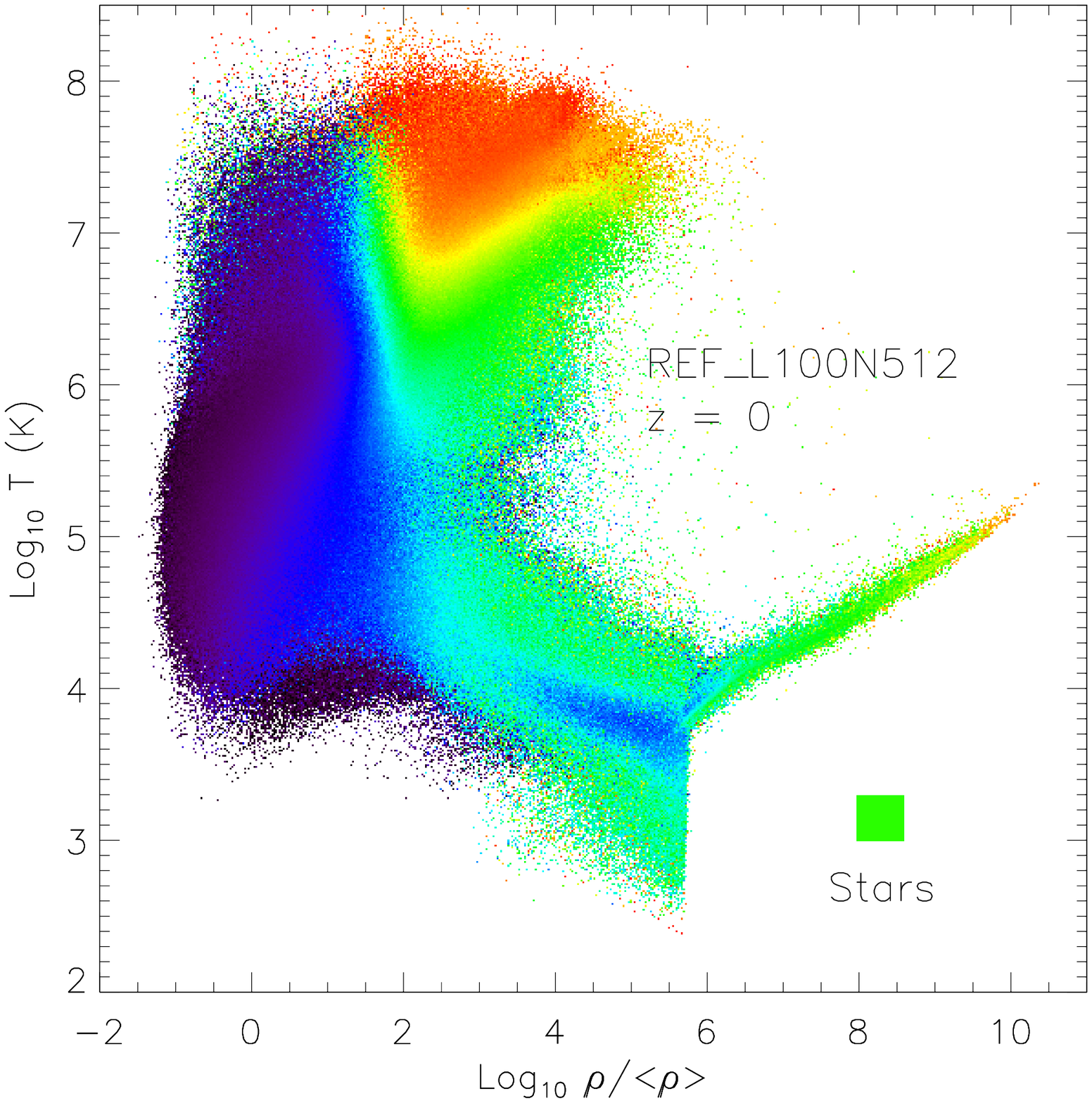}
\includegraphics[width=52.5mm]{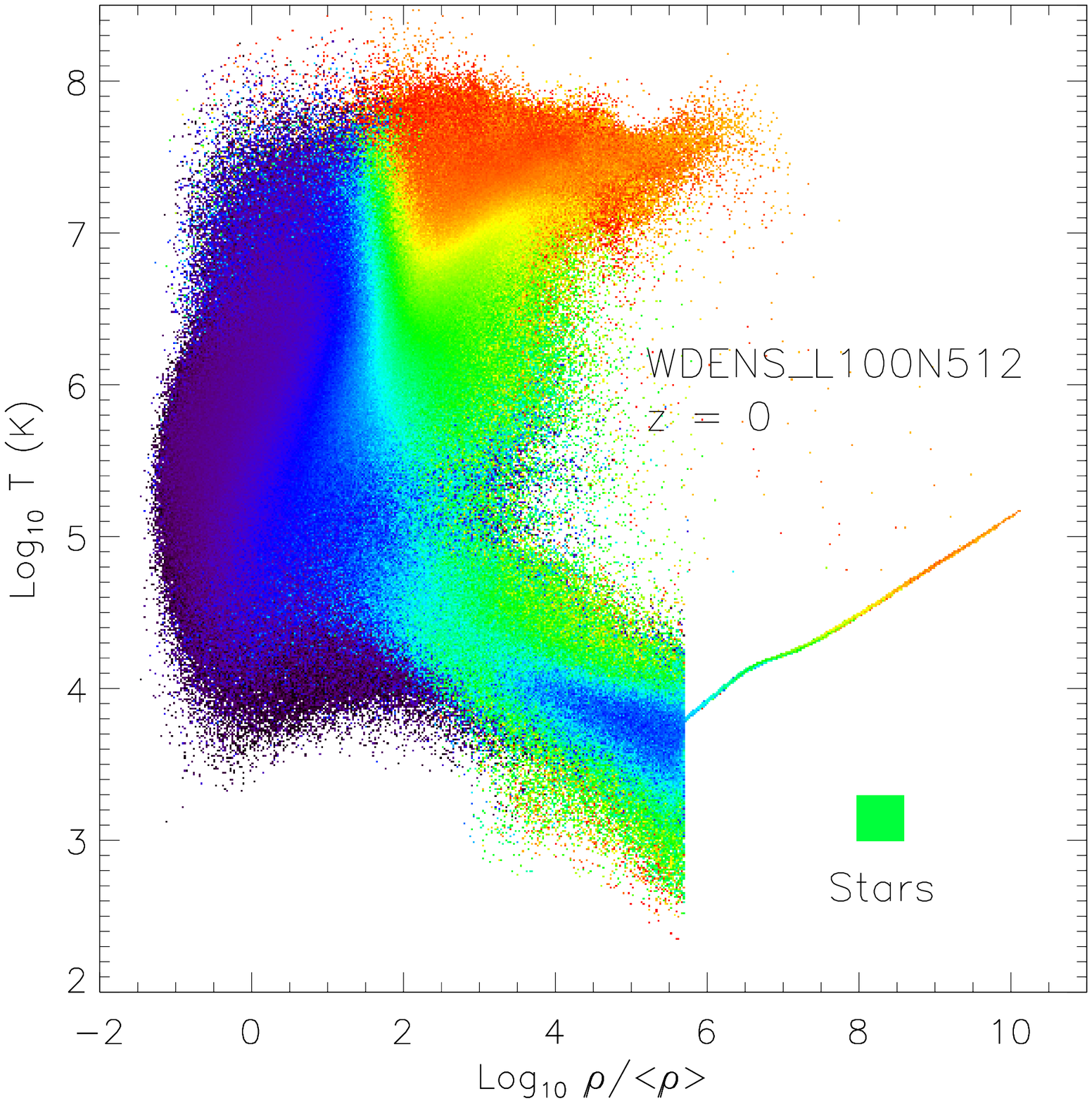}
\includegraphics[width=52.5mm]{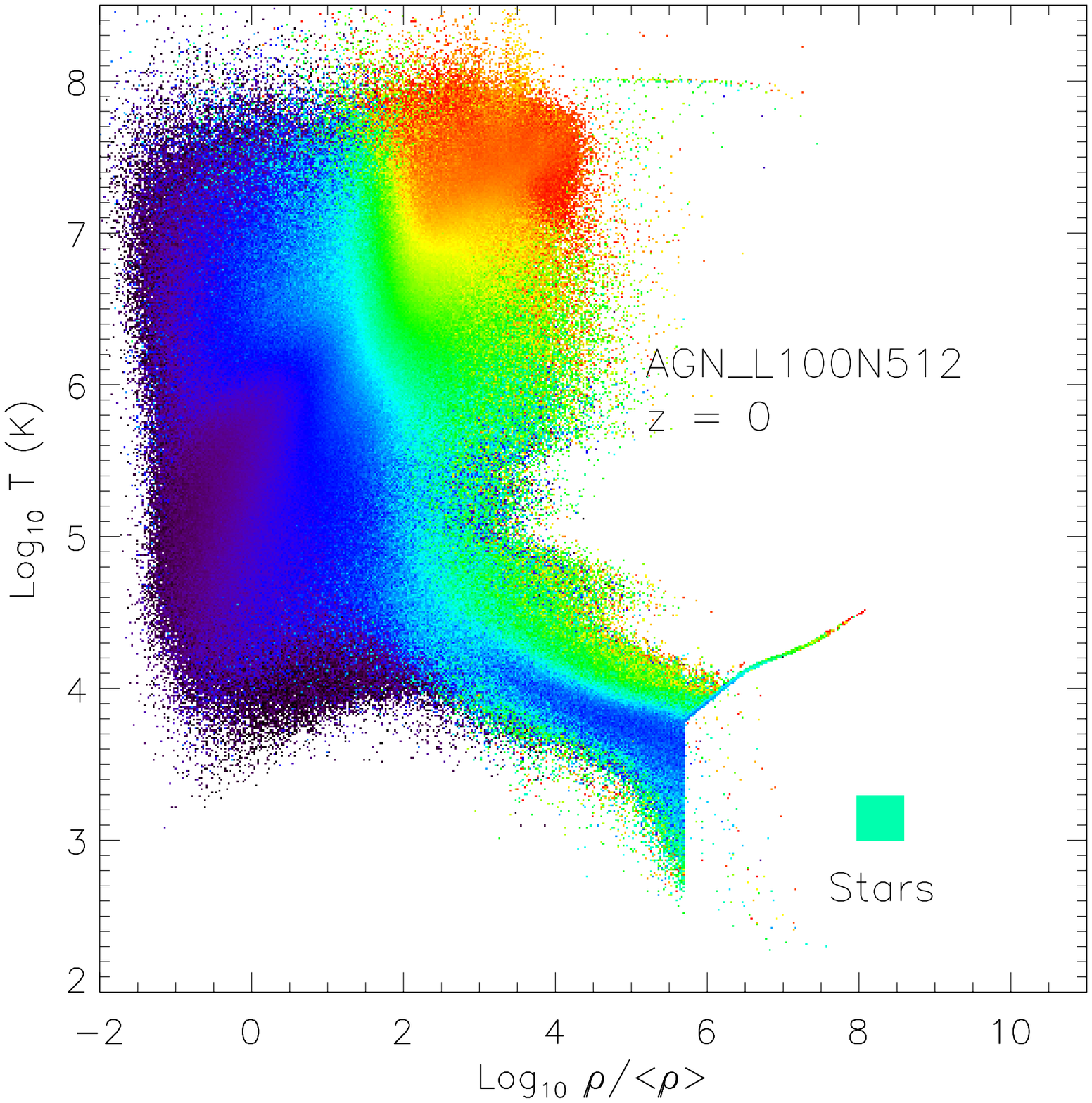}
\includegraphics[width=10.5mm]{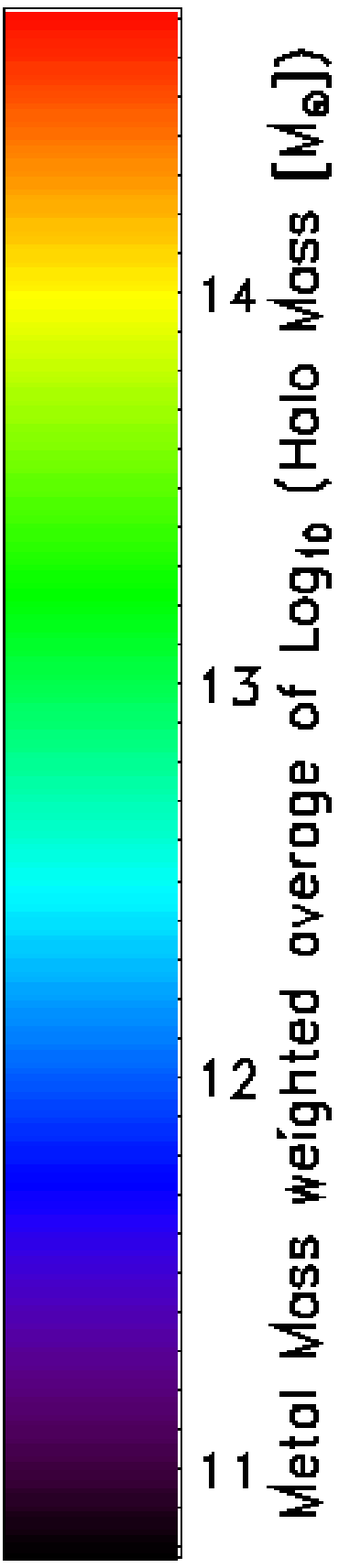}
\caption{As Fig.~\ref{fig-hmass} but for simulations \textit{REF\_L100N512} (left panel), \textit{WDENS\_L100N512} (middle panel), 
  and \textit{AGN\_L100N512} (right panel), all at $z=0$. Model \textit{WDENS} uses an implementation of winds driven by supernovae that is more efficient in massive haloes than the prescription used in model \textit{REF}. Model \textit{AGN} uses the same recipe for supernova feedback as model \textit{REF}, but also includes feedback from AGN. While there are some small differences, the trends are very similar for the three models. \label{fig-hmassL100}}
\end{figure*}

Now that we have a general impression of when the cosmic gas was
enriched, we would like to investigate what types of galaxies were
responsible for the enrichment. That is, do galaxies residing in certain halo masses preferentially enrich a typical phase of gas? We will address this question by determining the mass of the halo from which the metals were last ejected.

We identify haloes using a friends-of-friends group finder with a linking length of 0.2. This group finder is run on the dark matter particles in the simulation. Baryonic particles are associated with their nearest dark matter neighbour. We include only haloes with at least 100 dark matter particles, because we found that this limit results in excellent agreement with the \citet{Sheth2001} fit to the dark matter halo mass function. For model \emph{L100N512} this lower limit corresponds to a typical $z=0$ stellar mass of $\sim 10^{9.0}\,\Msun$ and this is also roughly the value below which the stellar mass function cuts off due to the limited resolution. For models \emph{L050N512} and \emph{L025N512}, however, the stellar mass already cuts off below about $10^{2.5}$ dark matter particles, which corresponds to stellar masses of $\sim 10^{8.3}\,\Msun$ for \emph{L050N512} at $z=0$ and $\sim 10^{7.7}\,\Msun$ for \emph{L025N512}. We therefore expect that we will underestimate the importance of the lowest mass haloes that we consider, particularly for these last two models.

To make the connection between metals and the haloes out of which
they were ejected, we perform the following calculation. For each enriched particle, we begin at $z = 0$ ($z = 2$ for the \textit{L025N512} simulation) and search for that particle (using its unique ID) in the friends-of-friends halo catalogue. If said particle is found, we note the total mass of the halo ($m_{\rm halo}$), and the corresponding redshift, $z_{\rm halo}$. For all particles not found in this output, we choose halo outputs at successively higher redshifts\footnote{Halo catalogues were generated at all simulation output redshifts. These are spaced at $\Delta z=0.125$ for $0\le z \le 0.5$, at $\Delta z=0.25$ for $0.5 \le z \le 4$, and at $\Delta z = 0.5$ for $4 \le z \le 9$. We also have outputs for $z = 10$, 12.5, 15, and 20.}, until we have
exhausted all of them. 
For every particle we thus have values for the  
temperature, density, and metal mass as well as $z_{\rm halo}$ and
$m_{\rm halo}$ (unless no corresponding halo is found in our outputs). 
Note that the temperature and density recorded are for 
$z = 0$ (or $z=2$), whereas $m_{\rm halo}$ is the mass of the host halo at redshift $z_{\rm halo}$.

If metals in the IGM were ejected from
the haloes in which they formed shortly after they were released by
stars, then the enrichment redshift, $z_Z$,
and $z_{\rm halo}$ should roughly agree. We would expect
the two to be similar, as both metal enrichment and outflows are
driven by massive stars, with $z_{\rm halo}$ somewhat smaller than
$z_Z$, simply because it takes some time for gas that has been ejected
to reach the virial radius and leave the halo. However, even if
enrichment and ejection happen nearly simultaneously, we would expect
differences of the order of the time scales corresponding to the time
intervals between snapshots. We have checked this and found that 
$z_{\rm  halo}$ and $z_z$ do indeed agree very well for almost all
temperatures and densities. This is true for both $z = 0$ and $z =
2$. An exception is hot, dense gas ($10^6 \K \la T \la 10^7 \K, \rho
\sim 10^2 \left < \rho \right >$) at $z=0$, for which $z_{\rm halo}$
is significantly smaller than $z_Z$. This may be because this gas
recently moved from the ICM across (our definition of) the halo
boundary, but was enriched at relatively high redshift.  

The colour scale in Fig.~\ref{fig-hmass} indicates the metal mass weighted mean $\log_{10} m_{\rm halo}$ as a function of the gas temperature and density for \textit{L050N512} at $z=0$ (left) and for \textit{L025N512} at $z=2$ (right). Particles that were never part of a halo containing at least 100 dark matter particles were excluded. These particles account for about 4 and 0.3 percent of the total metal mass in gas at $z=0$ and 2, respectively. Recall that gas with $\rho \ga 10^2 \left < \rho \right >$ typically resides in a halo.  

There is a strong gradient with density in both panels. The lower the density, the lower the mass of the objects that were responsible for the enrichment. The enrichment of the low-density IGM ($\rho < 10  \left < \rho \right >$) is dominated by the smallest, resolved haloes, particularly at $z=0$. This explains the relatively poor convergence that we found for these components in the previous section. Comparing with Fig.~\ref{fig-L050zwz}, we see that the low-density photo-ionised IGM ($\rho < 10  \left < \rho \right >$, $T < 10^{4.5} \K$) is nearly completely missing from Fig.~\ref{fig-hmass}. This gas phase has therefore only been enriched by haloes below our resolution limit and/or by intergalactic stars. The latter can occur when stars are ejected from galaxies via some dynamical process and then enrich the ambient IGM.

There is a positive gradient of $m_{\rm halo}$ with $T$ for high densities ($\rho \ga 10^2  \left < \rho \right >$). Since most of these particles reside in haloes, this merely reflects the fact that the virial temperature is an increasing function of mass.

\begin{figure*} 
\resizebox{\colwidth}{!}{\includegraphics{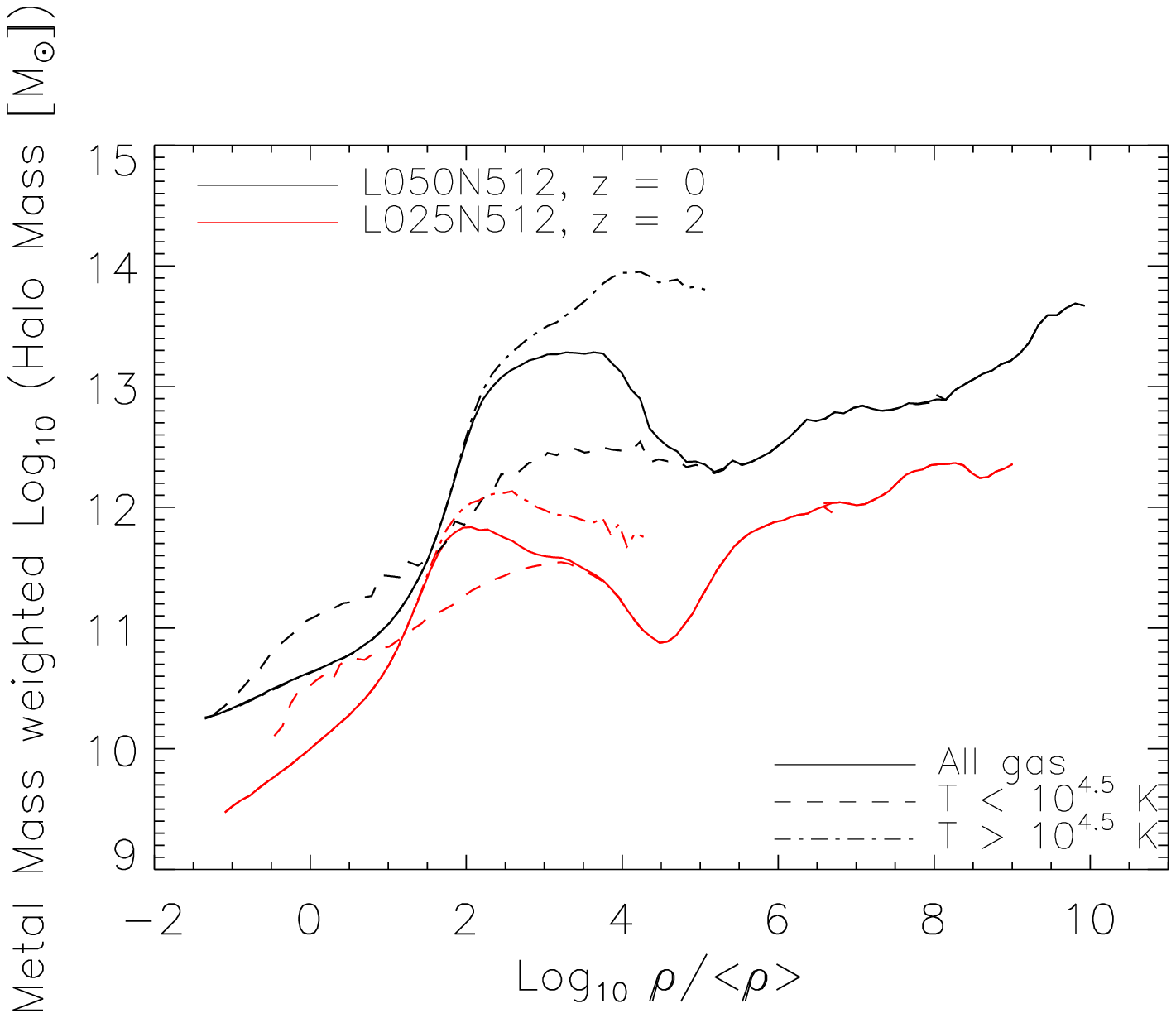}}
\resizebox{\colwidth}{!}{\includegraphics{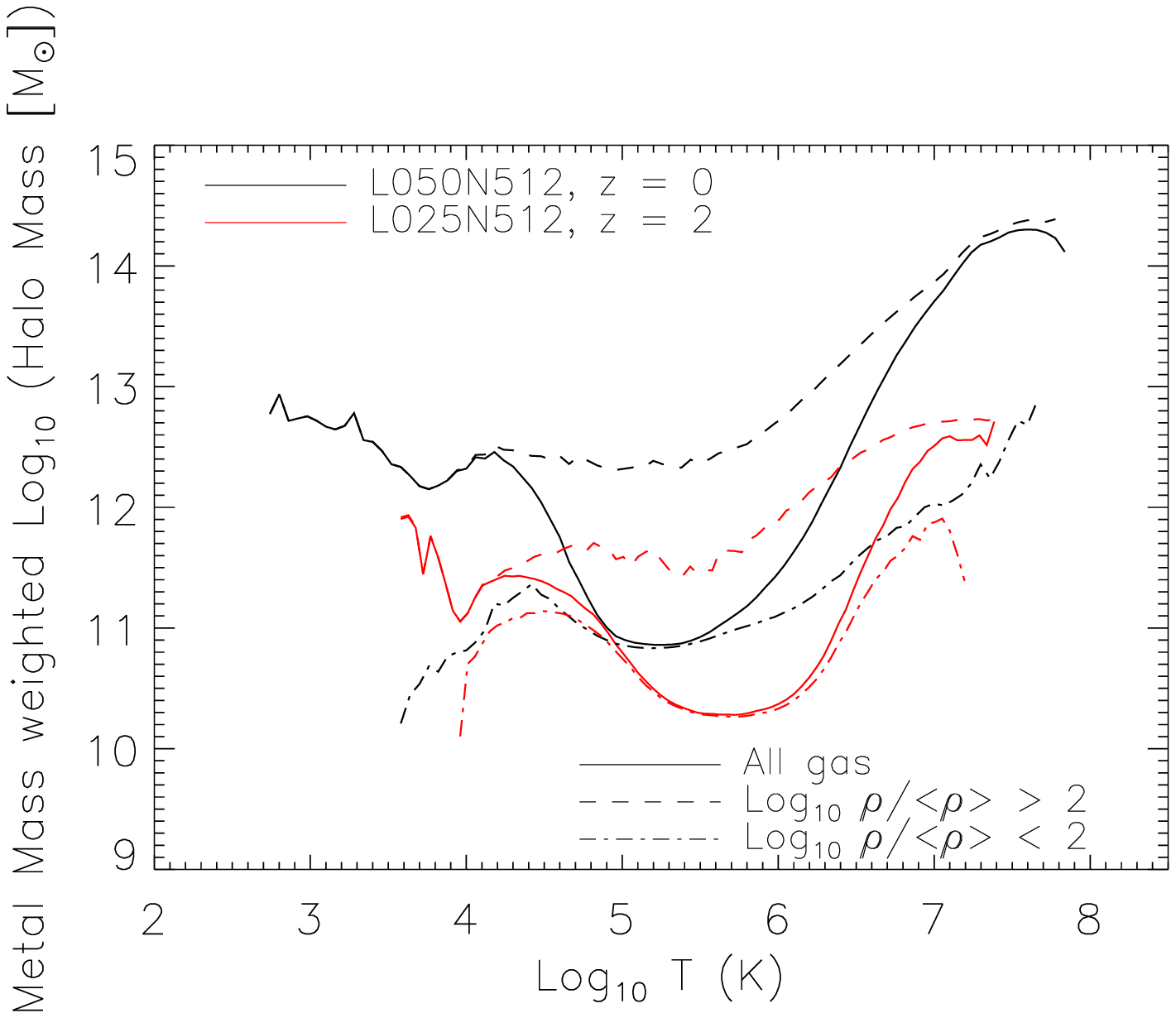}}\\
\caption{Metal mass weighted mean $\log_{10} m_{\rm halo}$, where $m_{\rm halo}$ is the total mass of the halo that last contained the metals, as a function of the gas density (left panel) or the gas temperature (right panel) at $z=0$ for model \textit{L050N512} (black) and at $z=2$ for \textit{L025N512} (red). The different line styles correspond to cuts in the temperature (left panel) or density (right panel) as indicated in the legends. The curves were cut when the number of particles per bin falls below 200. For both the diffuse IGM and the ISM, the typical halo mass that last contained the metals increases with the gas density.
Gas at intermediate temperatures ($T\sim 10^5 - 10^6\,\K$) typically received its metals from lower mass haloes, but this merely reflects the fact that metals at these temperatures reside mostly in lower density gas than metals at lower or higher temperatures, as can be seen by comparing the different line styles in the right panel.  \label{fig-hmasshist}}
\end{figure*}

Fig.~\ref{fig-hmassL100} compares the reference model to model \textit{WDENS}, which uses an implementation of galactic winds that makes them more efficient in high-mass haloes, and to model \textit{AGN}, which includes AGN feedback. While there are some small differences, the trends are identical. Thus, the results are robust to variations in the numerical implementation of galactic winds and to the inclusion of AGN feedback. Henceforth we will therefore only consider the reference model. 

\begin{figure*} 
\includegraphics[width=168mm]{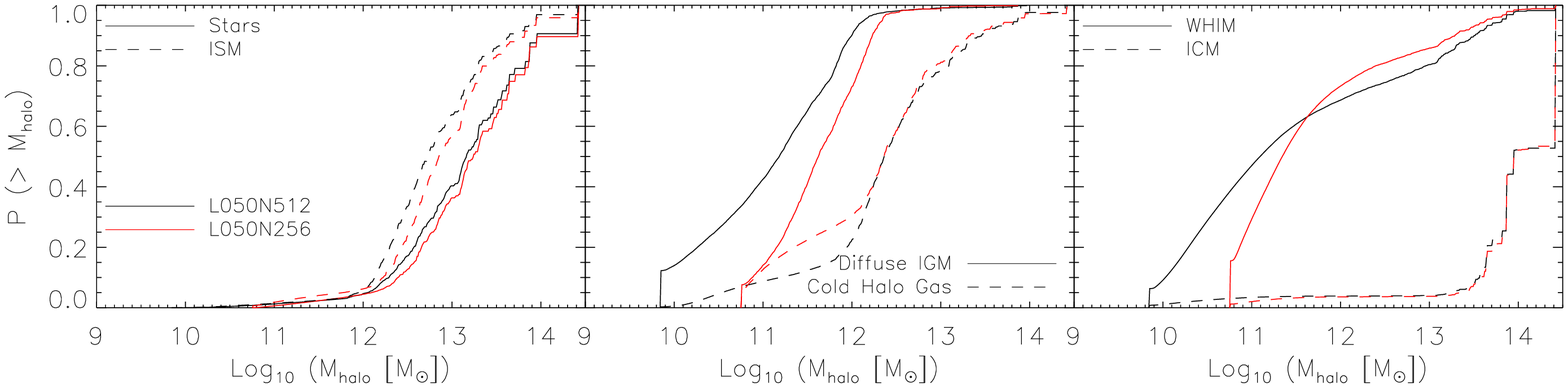}\\
\includegraphics[width=168mm]{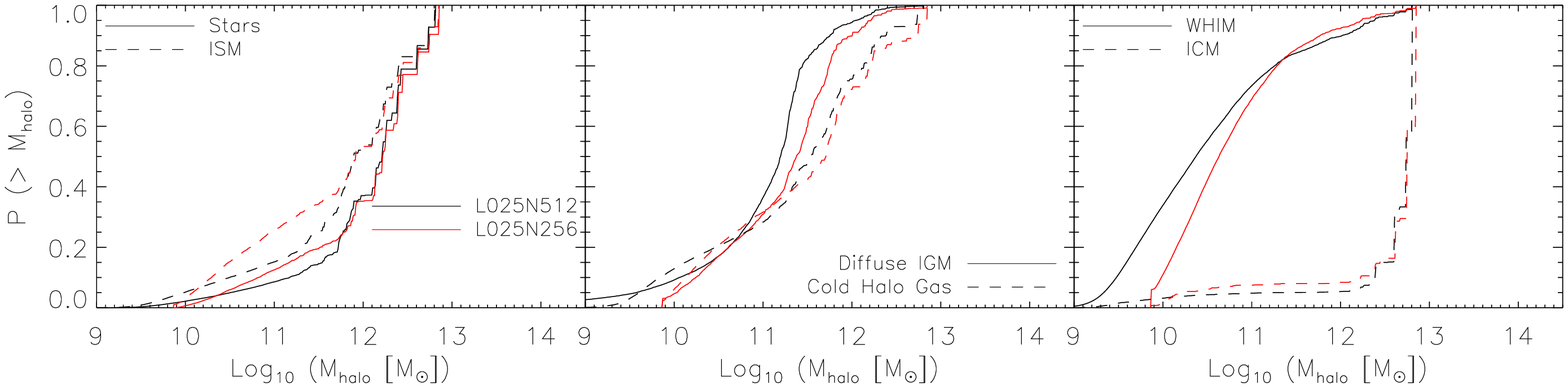}
\caption{Cumulative probability distribution of the halo mass that last contained the metals, weighted by metal mass, at $z=0$ (top row) and $z=2$ (bottom row) for stars and for the different gas phases defined in Fig.~\protect\ref{fig-smZcont}. The different colours correspond to different numerical resolutions. For metals that have never resided in a halo with more then 100 dark matter particles we set $m_{\rm halo}$ to the minimum resolved halo mass. The results are close to converged with respect to resolution except for the diffuse IGM and the WHIM. For these phases even our highest resolution runs may overestimate the halo mass. A wide range of halo masses contribute to the enrichment. \label{fig-enrichhalo}}
\end{figure*}

Comparison of the left panel of Fig.~\ref{fig-hmassL100} to the left panel of Fig.~\ref{fig-hmass}, which uses a mass
(spatial) resolution that is eight (two) times higher, we see that the
trends are also robust with respect to the numerical resolution. However, there are clear quantitative differences (note that the colour scales differ), with the higher resolution simulation giving lower halo masses, particularly for masses near the resolution limit. 

Fig.~\ref{fig-hmasshist} shows the metal mass weighted mean $\log_{10}$ of the halo mass that last contained the metals as a function of density (left panel) and temperature (right panel) for \textit{L050N512} at $z=0$ (black) and for \textit{L025N512} at $z=2$ (red). This plot therefore shows projections of Fig.~\ref{fig-hmass}, weighted by the metal mass in each pixel of that figure. The solid curves show the results for all the gas, while the other line styles correspond to cuts in the temperature (left) or density (right). Note that due to the finite resolution of the simulation, we may strongly overestimate the typical halo mass of origin when it is predicted to be of the order the minimum resolved halo mass. 

The solid curves in the left panel follow the dot-dashed curves at low densities ($\rho \la 10^2 \left <\rho\right >$) because most of the metals residing in diffuse gas are hot ($T> 10^{4.5}\,\K$). Conversely, for densities of order or greater than the star formation threshold, most of the metals are cold and the solid curves follow the dashed curves. We see similar behaviour in the right panel.

The typical halo mass out of which the metals residing in the diffuse IGM were ejected increases strongly with the gas density. As the density increases above that typical of collapsed objects ($\rho \sim 10^2 \left < \rho \right >$), $m_{\rm halo}$ firsts declines, because cold metals begin to account for a substantial fraction of the metal mass and for lower mass haloes a greater fraction of the gas is in clumps. Beyond the star formation threshold the mean halo mass increases again as more massive haloes sample higher pressure and thus, given our imposed equation of state for star-forming gas, higher density gas. The right panel shows that, for a fixed temperature, metals in higher density gas originate in higher mass haloes. The typical halo mass that last contained the metals reaches a minimum at intermediate temperatures ($T\sim 10^5 - 10^6\,\K$), because at these temperatures most of the metals reside at low-densities ($\rho < 10^2\left < \rho \right >$), while high-density gas dominates the metal content at both low ($T \la 10^4\,\K$) and high ($T\ga 10^7\,\K$) temperatures.

In order to gain insight into the distribution of halo masses that contribute to the enrichment of a particular gas phase, we show the cumulative probability distribution function of $m_{\rm halo}$ in Fig.~\ref{fig-enrichhalo}. Phrased differently, the figure shows the fraction of the metal mass that last resided in halo masses exceeding those plotted along the $x$-axis. The different columns and line styles correspond to the different regions in $T-\rho$ space shown in Fig.~\ref{fig-smZcont} and the different colours show the results for different resolutions. The top and bottom rows are for $z=0$ and $z=2$, respectively. 

For metals that never resided in a resolved halo we set $m_{\rm halo}$ to the dark matter mass of the smallest resolved haloes. The fraction of the metal mass ejected by unresolved haloes is greatest for the diffuse IGM at $z=0$, for which it is about 12 percent.  
Comparing the two resolutions, we see that the results for the low-density phases, i.e.\ the diffuse IGM and the WHIM, are not converged. The situation is worst for the diffuse IGM at $z=0$. 
Note that within these phases, convergence will be worse at lower densities as we have already demonstrated that the mean halo mass of origin decreases with decreasing density. Fig.~\ref{fig-enrichhalo} clearly shows that a wide range of halo masses contributes to the enrichment of most gas phases. More than half of the metal mass in the diffuse IGM and the WHIM originated from haloes with total mass $M<10^{11}\,\Msun$, which corresponds to stellar masses smaller than $10^9\,\Msun$. For the WHIM at $z=2$ this fraction is about 80 \%. As we noted in \S\ref{sec-when}, if we separate the WHIM at $\rho = 100~\langle \rho \rangle$ (not shown), then the low-density component behaves similarly to the full WHIM, while the high-density component is more similar to the ICM.

\section{What enriched the IGM and when?}
\label{sec-conn}

\begin{figure*}
\includegraphics[width=84mm]{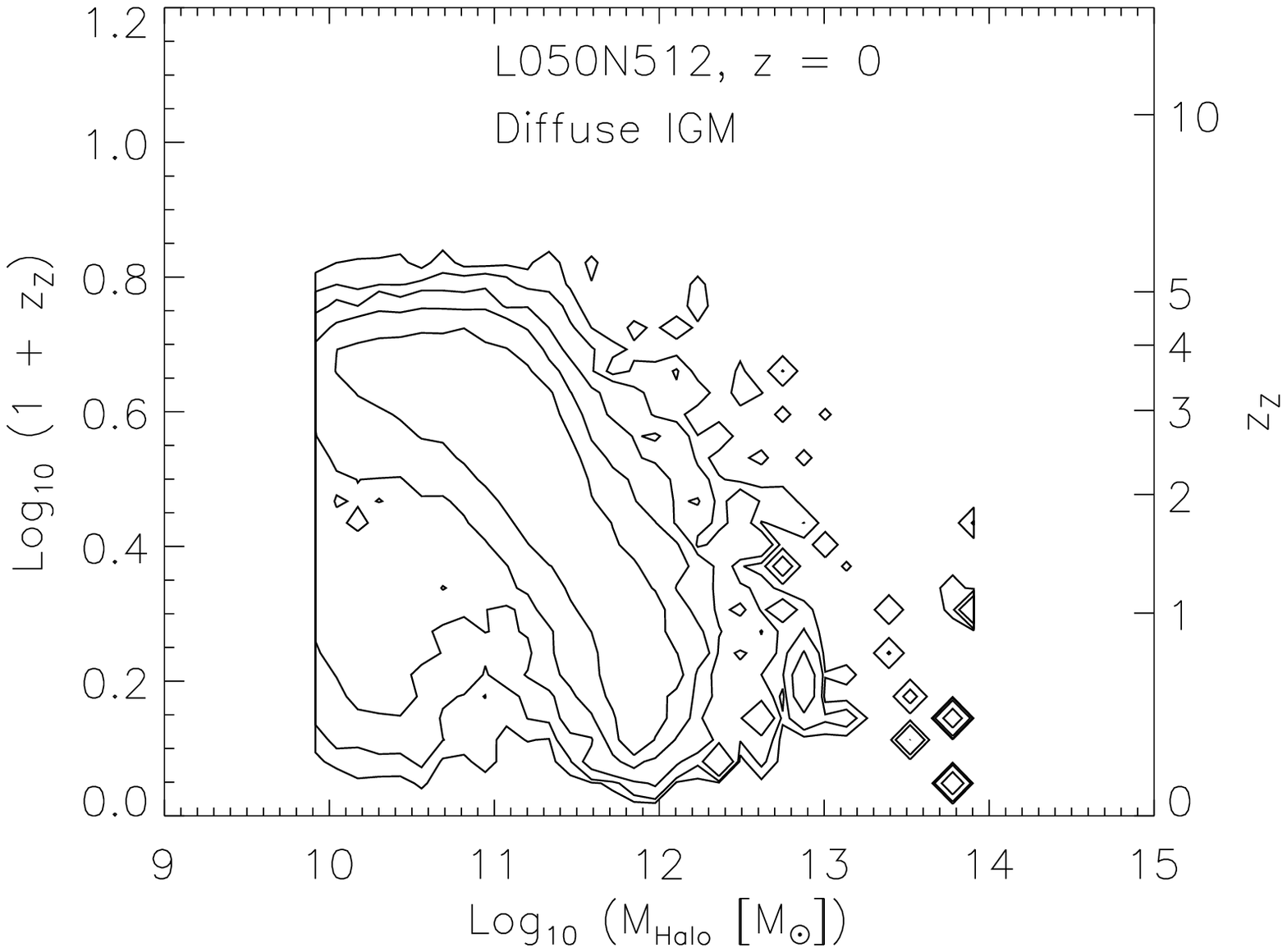}
\includegraphics[width=84mm]{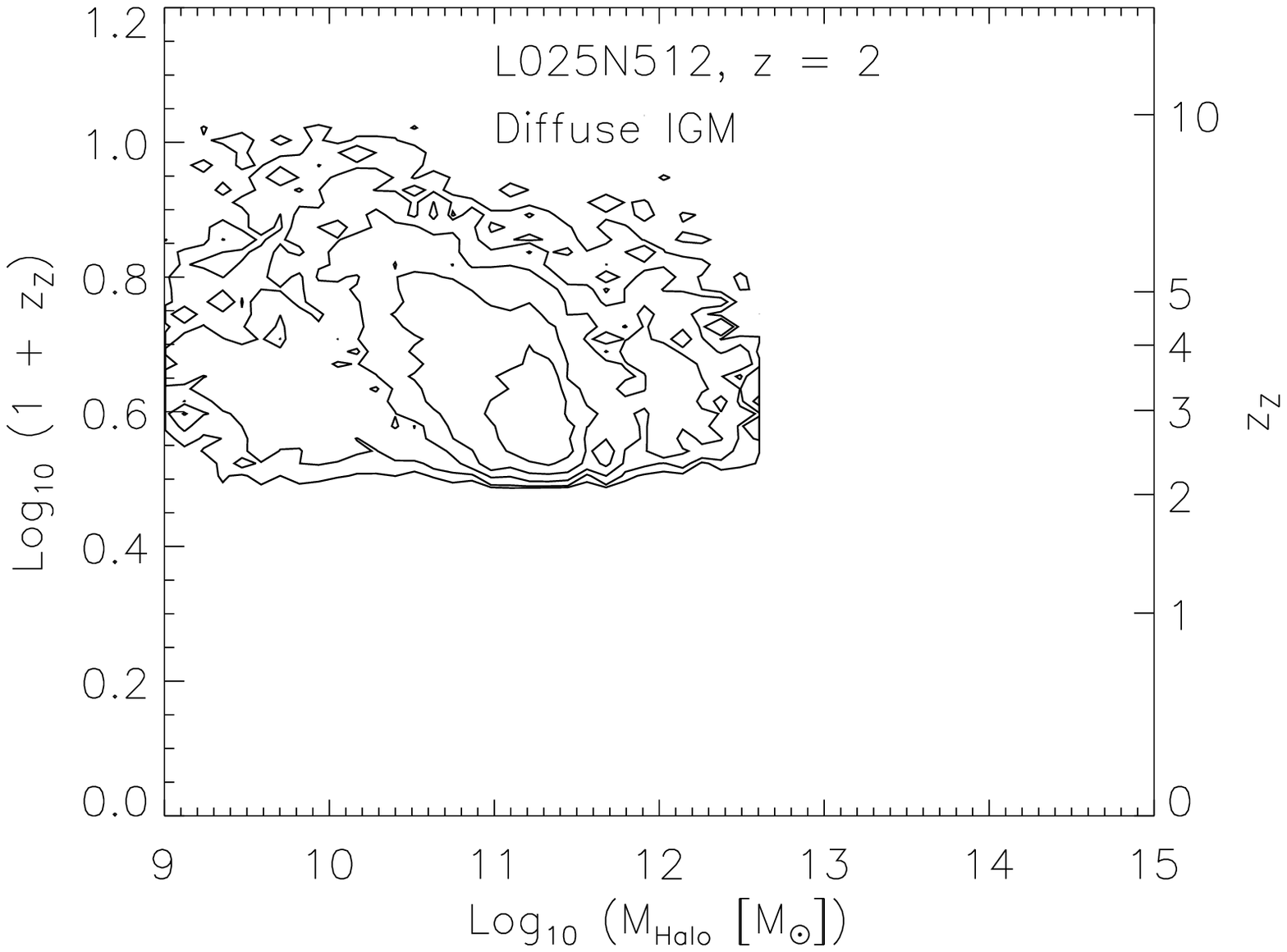}\\
\includegraphics[width=84mm]{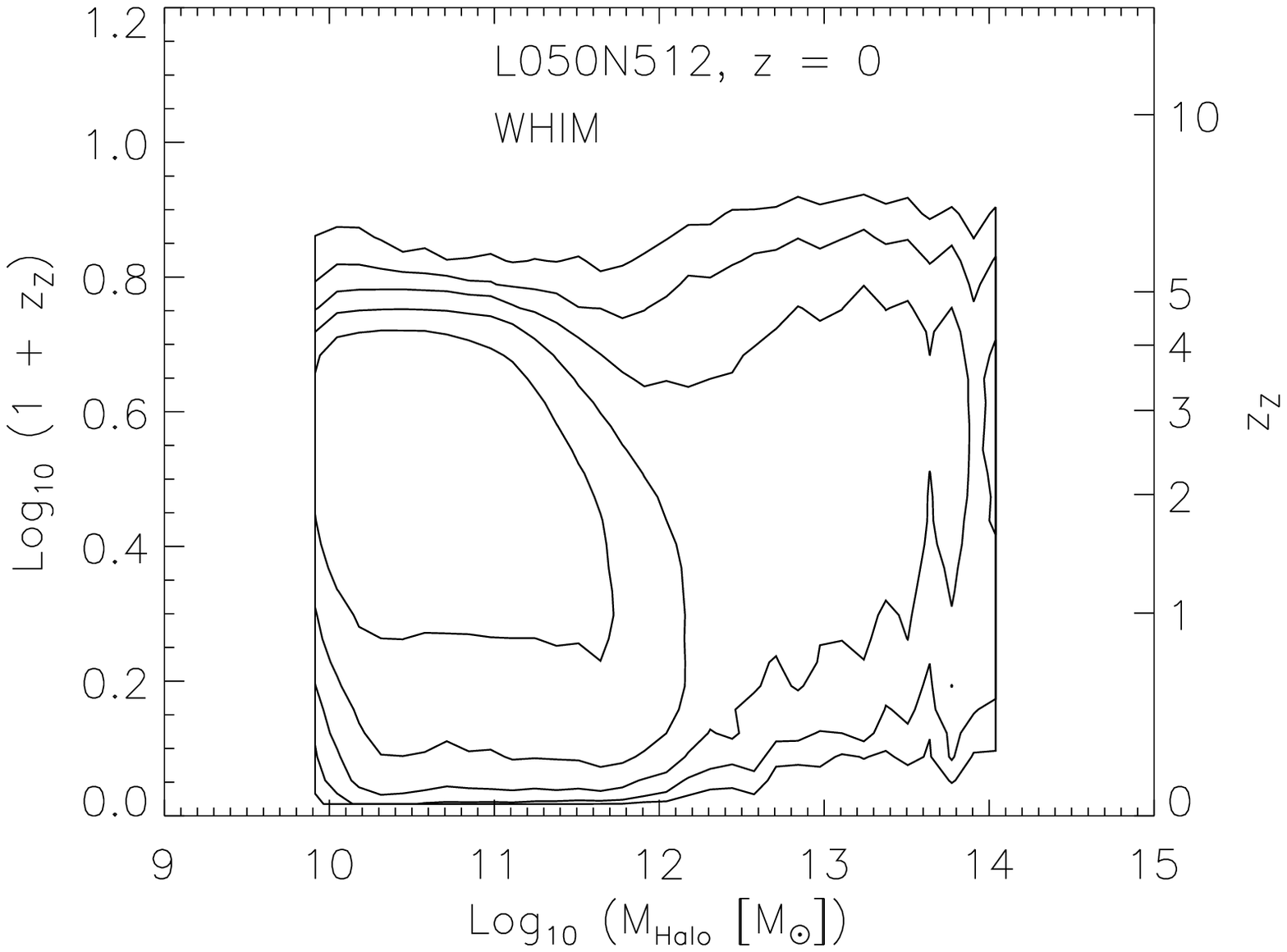}
\includegraphics[width=84mm]{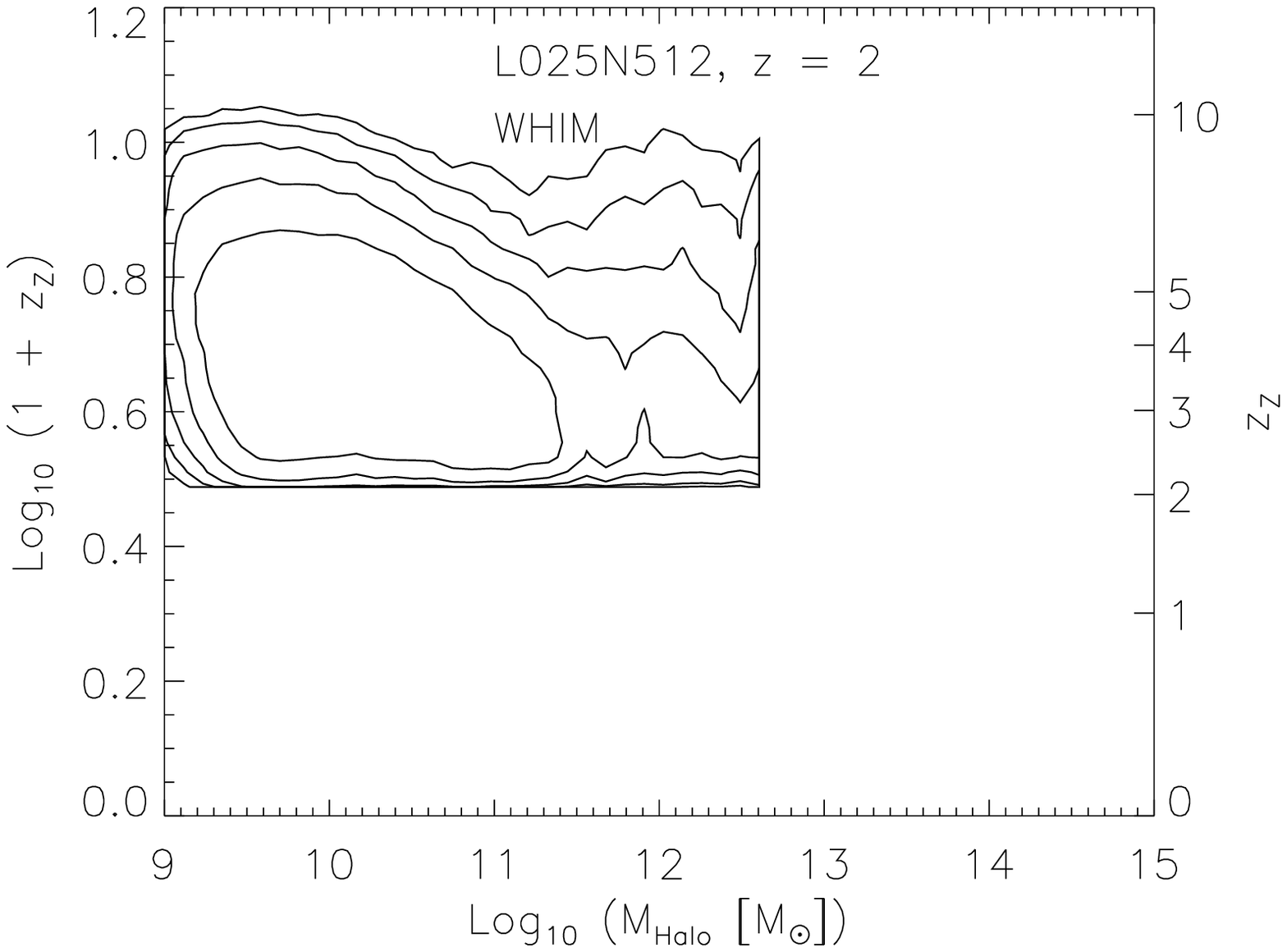}
\caption{Distribution of metal mass as a function of $m_{\rm halo}$, the metal mass weighted mean mass of the last resolved halo that contained the metals, and $z_Z$, the metal mass weighted mean enrichment redshift. Results are shown for the diffuse IGM (top row) and the WHIM (bottom row) at $z=0$ for \textit{L050N512} (left column) and at $z=2$ for \textit{L025N512} (right column). The contours are logarithmically spaced by 0.5 dex. Metals that are currently in a halo and metals that have never been part of a resolved halo were both excluded. Older metals were typically ejected by lower mass haloes, particularly if they end up in the diffuse IGM. The WHIM was enriched by a wide range of halo masses. \label{fig-masszwz}}
\end{figure*}

We conclude by simultaneously addressing the two questions that we investigated in the two previous sections: When was the IGM enriched and by what? Fig.~\ref{fig-masszwz} shows the metal mass distribution as a function of both $z_Z$ and $m_{\rm halo}$ for model \textit{L050N512} at $z=0$ (left column) and for model \textit{L025N512} at $z=2$ (right column). The top and bottom rows show the results for metals residing in the diffuse IGM and the WHIM, respectively. Metals that are currently part of a halo were excluded (note that this corresponds to a large fraction of the metals for the case of the WHIM). As discussed before, the finite resolution of our simulations will cause us to underestimate the importance of haloes with masses near and below the minimum halo mass that we consider. 

Focusing first on the diffuse IGM at $z=0$ (the top-left panel of Fig.~\ref{fig-masszwz}), we see that there is a strong anti-correlation between the enrichment redshift and the mass of the halo from which the metals were ejected. Early enrichment is more strongly dominated by lower mass galaxies. For metals residing in the diffuse IGM at $z=2$ (top-right panel) this is also the case, but the enrichment seems to have been dominated by objects with $m_{\rm halo}\sim 10^{11}\,\Msun$ (which corresponds to a stellar mass $m_\ast\sim 10^9\,\Msun$), which is about two orders of magnitude greater than the minimum halo mass we consider. The smaller role for the lowest mass haloes at $z=2$ is also apparent from a comparison of the solid curves in the middle panels of Fig.~\ref{fig-enrichhalo}. It appears that the resolution of the \textit{L025N512} simulation is nearly sufficient to model the enrichment of the diffuse IGM at $z=2$. 

We must keep in mind, however, that the diffuse IGM is not a single phase. We have already demonstrated that $z_Z$ and $m_{\rm halo}$ are strongly increasing resp.\ decreasing functions of the gas density (see Figs.~\ref{fig-zwzcontour} and \ref{fig-hmasshist}). For example, for the diffuse IGM at $z=2$ the metal mass weighted mean $\log_{10}m_{\rm halo}$ decreases from $\approx 12$ for an overdensity of $10^2$ to $\approx 10$ (i.e.\ too close to the resolution limit to be converged) at the mean density. 

It is likely that a significant fraction of the metal mass that resides at the high-density end of the diffuse IGM at $z=2$, which is typically ejected by more massive haloes, will have fallen back into haloes by $z=0$ and will therefore no longer be part of the diffuse IGM. This effect can help explain why $m_{\rm halo}$ is smaller at $z=0$ than at $z=2$ and why there is an anti-correlation between $z_Z$ and $m_{\rm halo}$. Evolution of the halo mass function is likely also an important part of the explanation. At higher redshifts the masses of the haloes producing metals are typically lower. 

The enrichment history of the WHIM (bottom row of Fig.~\ref{fig-masszwz}) shows some interesting differences from that of the diffuse IGM. At $z=0$ the anti-correlation between $z_Z$ and $m_{\rm halo}$ is much weaker. Relatively recent enrichment by low-mass haloes is more important for the WHIM. Presumably, metals that are ejected recently are shock-heated by the winds that carry them and many of them may not have had sufficient time to cool down to the temperatures of the diffuse IGM. High-mass haloes are also more important for the WHIM. This may largely be because high-mass haloes tend to be embedded in gas that is heated by gravitational accretion shocks and is therefore sufficiently hot to be considered part of the WHIM.

The results for the other OWLS models considered here are very similar (not shown), although higher mass haloes are slightly more important for \textit{WDENS} and \textit{AGN}, presumably because in these models feedback is more efficient in such objects.

\section{Conclusions}
\label{sec-summary}

We have used cosmological, hydrodynamical simulations that include radiative cooling, star formation, stellar evolution and supernova feedback, and, in some cases, AGN feedback, to investigate the metal enrichment history of the Universe. Specifically, we took advantage of the Lagrangian nature of our simulations to investigate when gas that ends up in a particular gas phase at redshift $z=0$ or $z=2$ received its metals and what the masses were of the haloes that last contained these metals. 

We considered different physical models, all taken from the OWLS
project, including simulations with a different implementation of
supernova feedback and a model that includes AGN feedback. While there
are some small differences, such as slightly higher enrichment
redshifts for models with more efficient feedback in massive galaxies,
the main trends are strikingly similar for all models. We cannot rule out 
that some of our conclusions may depend on our prescriptions for 
feedback or that they are specific to SPH. However, the fact that three 
drastically different feedback models give such similar results suggests 
that our conclusions are robust. 

The time since the enrichment varies most strongly with the gas density. Metals in lower density gas are typically much older than metals in high-density gas, a trend that extends over ten orders of magnitude in gas density. At least half of the metals that reside in the diffuse IGM at $z=0$ ($z=2$) were ejected above redshift two (three). 

The enrichment redshift also varies with the temperature of the gas, but this mostly reflects the fact that the typical gas density of enriched gas varies with its temperature. For gas that is dense enough to be able to cool, the time since enrichment correlates well with the cooling time. Gas with shorter cooling times received its metals more recently. 

The typical mass of the haloes from which the metals residing in the IGM were ejected, increases rapidly with the gas density. At least half of the metal mass was ejected by haloes with total masses less than $10^{11}\,\Msun$, which corresponds to stellar masses smaller than $10^9\,\Msun$. For the low-redshift IGM the mass of the dominant IGM polluters may be substantially smaller than this because our predictions for present day low-density gas are not close to converged with respect to the numerical resolution. 

The age of the metals ending up in the diffuse IGM is strongly anti-correlated with the mass of the haloes from which they were ejected. In other words, older metals were typically ejected by lower mass haloes. This anti-correlation is less strong for the hotter part of the IGM (the WHIM), for which recent enrichment by low-mass haloes is more important. While massive haloes ($> 10^{12}\,\Msun$) are unimportant for the enrichment of the diffuse IGM, they do contribute to the pollution of the warm-hot IGM at low redshift.  

Our main and strongest conclusion is that metals residing in lower density gas were typically ejected earlier and by lower mass haloes. This suggests that travel-time is a limiting factor for the enrichment of the IGM, as proposed by \citet{Aguirre2001z3}. Metals that have been ejected recently simply have not had sufficient time to reach the low-density gas far from galaxies. The anti-correlation between metal age and halo mass would then follow because typical galaxy masses are lower at higher redshifts. Another mechanism that could explain the trends is fall back onto galaxies. If metals ejected by lower mass galaxies are less likely to fall back \cite[e.g.][]{Oppenheimer2010} or if they typically leave the haloes with larger velocities, then we would expect lower mass galaxies to be more important for gas that is further away. Finally, low-mass galaxies are more weakly clustered than high-mass galaxies, thus low-density gas is more accessible to low-mass galaxies.

It is not immediately obvious how one could compare this result with observations. One way would be to consider abundance ratios of elements in various phases 
as a function of density and redshift in an attempt to find a signature that could be searched for in high-redshift observations. However, uncertainties in the nucleosynthetic yields and the type Ia supernova rates \citep[see][]{Wiersma2009b}, as well as in the ionization corrections, may make such a comparison troublesome. Correlations between absorbers and the distances to different types of galaxies may also help shed light on the enrichment mechanism \citep[e.g.][]{Steidel2010,Chen2009,Wilman2007,Stocke2006}, although is is important to keep in mind that the metals that are observed to be near a galaxy may have been injected by its lower-mass progenitors \citep{Porciani2005,Scannapieco2005}. In fact, our results would imply that this is likely to be the case for metals in low-density gas. We leave an analysis of abundance ratios and other observational comparisons to a future work.

The importance of low-mass galaxies is a challenge for simulations of the enrichment of the IGM, as it implies the need for high-resolution in large volumes. On the other hand, our results suggest that the low-density IGM provides us with an exciting opportunity to study the consequences of feedback in low-mass galaxies, a key but poorly understood ingredient of models of galaxy formation, and that it provides us with a fossil record of galaxy formation in the high-redshift Universe.

\section*{Acknowledgements}
We are very grateful to Volker Springel for invaluable help with the simulations and for a careful reading of the manuscript. We would also like to thank Ben Oppenheimer for a careful reading of the manuscript.
The simulations presented here were run on Stella, the LOFAR Blue
Gene/L system in Groningen, on the Cosmology Machine at the Institute
for Computational Cosmology in Durham as part of the Virgo
Consortium research programme and on Darwin in Cambridge. This
work was sponsored by National Computing Facilities Foundation
(NCF) for the use of supercomputer facilities, with financial support
from the Netherlands Organization for Scientific Research (NWO), Marie Curie Excellence Grant MEXT-CT-2004-014112, an NWO VIDI grant, NSF grants AST-0507117 and AST-0908910, and DFG Priority Program 1177.

\bibliographystyle{mn2e} 
\bibliography{ms}

\end{document}